\documentclass{jaa}
\usepackage{natbib}
\usepackage{graphicx}
\usepackage{enumitem,amssymb,amsmath}
\usepackage{ragged2e}
\usepackage{comment}
\usepackage[numbib]{tocbibind}
\usepackage{color}
\usepackage{soul}

\newcommand{\hii}{H {\sc ii}}
\newcommand{\HI}{H {\sc i}}

\newcommand{\CII}{[C {\sc ii}]}
\newcommand{\OI}{[O {\sc i}]}
\newcommand\micron{\mbox{$\mu$m}}
\settocbibname{References}
\bibpunct{(}{)}{,}{a}{}{,}
\pgrange{1--}
\lp{28}

\begin{document}\sloppy

\label{firstpage}
\title{Research on the Interstellar Medium and Star Formation in the Galaxy: An Indian Perspective}

\author{Bhaswati Mookerjea\textsuperscript{1}, 
Maheswar G.\textsuperscript{2}, 
Kinsuk Acharyya\textsuperscript{3},
Tapas Baug\textsuperscript{4}, 
Prasun Datta\textsuperscript{5}, 
Jessy Jose\textsuperscript{6},
D. K. Ojha\textsuperscript{1}, 
Jagadheep D. Pandian\textsuperscript{7}, 
Nirupam Roy\textsuperscript{8}, 
Manash Samal\textsuperscript{9}, 
Saurabh Sharma\textsuperscript{10},
Archana Soam\textsuperscript{2}, 
Sarita Vig\textsuperscript{7},
Ankan Das\textsuperscript{11},
Lokesh Dewangan\textsuperscript{9},
Somnath Dutta\textsuperscript{12},
C. Eswariah\textsuperscript{6},
Liton Majumdar \textsuperscript{13,14},
Kshitiz Kumar Mallick\textsuperscript{15},
Soumen Mondal\textsuperscript{4},
Joe P. Ninan\textsuperscript{1},
Neelam Panwar\textsuperscript{10},
Amit Pathak\textsuperscript{16},  
Shantanu Rastogi\textsuperscript{17},  
Dipen Sahu\textsuperscript{10},
Anandmayee Tej\textsuperscript{7},
Veena. V.S\textsuperscript{18}}
\affilOne{\textsuperscript{1}Department of Astronomy \& Astrophysics, Tata Institute of Fundamental Research, Mumbai 400005, India, }
\affilTwo{\textsuperscript{2}Indian Institute of Astrophysics (IIA), Sarjapur Road, Koramangala, Bangalore 560034, India, }
\affilThree{\textsuperscript{3}Planetary Sciences Division, Physical Research Laboratory, Ahmedabad 380009, India, }
\affilFour{\textsuperscript{4}Satyendra Nath Bose National Centre for Basic Sciences, Salt Lake, Kolkata 700 106, India, }
\affilFive{\textsuperscript{5}Department of Physics, IIT (BHU), Varanasi 221005, India, }
\affilSix{\textsuperscript{6}Indian Institute of Science Education and Research (IISER) Tirupati,  Tirupati 517507, India, }
\affilSeven{\textsuperscript{7}Indian Institute of Space Science and Technology (IIST), Trivandrum 695 547, India, }
\affilEight{\textsuperscript{8}Department of Physics, Indian Institute of Science, Bangalore 560012, India, }
\affilNine{\textsuperscript{9}Astronomy \& Astrophysics Division, Physical Research Laboratory, Ahmedabad, 380009, India, }
\affilTen{\textsuperscript{10}Aryabhatta Research Institute of Observational SciencES (ARIES), Manora Peak, Nainital 263001, India,}
\affilEleven{\textsuperscript{11}Institute of Astronomy Space and Earth Science, CIT Road, Scheme 7m, Kolkata 700054, India, }
\affilTwelve{\textsuperscript{12}Institute of Astronomy and Astrophysics, Academia Sinica, Taipei 10617, Taiwan, }
\affilThirteen{\textsuperscript{13}National Institute of Science Education and Research, Bhubaneswar, Jatni, Khurda, Odisha, India, }
\affilFourteen{\textsuperscript{14}Homi Bhabha National Institute, Training School Complex, Anushaktinagar, Mumbai 400094, }
\affilFifteen{\textsuperscript{15}National Astronomical Observatory of Japan, 2-21-1 Osawa, Mitaka, Tokyo 181-8588, Japan, }
\affilSixteen{\textsuperscript{16}Department of Physics, Institute of Science, Banaras Hindu University, Varanasi-221005, India, }
\affilSeventeen{\textsuperscript{17}Department of Physics, D.D.U. Gorakhpur University, Gorakhpur,  273009, India, }
\affilEighteen{\textsuperscript{18}Max Planck Institut f\"ur Radioastronomie, Auf dem H\"ugel 69, D-53121 Bonn, Germany\\}
\twocolumn[{
\maketitle
\corres{bhaswati@tifr.res.in, maheswar.g@iiap.res.in}
\msinfo{}{}


\begin{abstract}
Although the star formation process has been studied for decades, many
important aspects of the physics involved remain unsolved. Recent
advancement of instrumentation in the infrared, far-infrared and
sub-millimetre wavelength regimes have contributed to a significantly
improved understanding of processes in the interstellar medium (ISM)
leading to star formation.  The future of research on the ISM and star
formation looks exciting with instruments like the JWST, ALMA, etc.,
already contributing to the topic by gathering high-resolution
high-sensitivity data and with several larger ground- and space-bound
facilities either being planned or constructed. India has a sizable
number of astronomers engaged in research on topics related to the ISM
and star formation. In this white paper invited by the Astronomical
Society of India to prepare a vision document for Indian astronomy, we
review the Indian contributions to the global understanding of the star
formation process and suggest areas that require focused efforts both in
creating observing facilities and in theoretical front in India, in
order to improve the impact of our research in the coming decades.
\end{abstract}

}]

\section{Introduction}

The interstellar medium (ISM), the matter, radiation and magnetic field (B-field) in the vast space between the stars, is a major constituent of the galaxies. For a Milky Way like galaxy, the ISM mass \citep[$\sim 10^{10}$\,M$_\odot$;][]{misiriotis2006} is about $10 - 15\%$ of the total stellar mass of the galaxy \citep[$\sim 6\times10^{10}$\,M$_\odot$;][]{licquia2015}. In the context of the star formation studies, the molecular clouds can be viewed as the fundamental ingredients of galaxies, since they are the channels that transform the diffuse atomic interstellar gas into stars. Given that most of the volume in the Milky Way is filled by atomic gas several times more diffuse than the molecular gas, understanding the transition to dense molecular gas and evolution of molecular clouds is an important first step in the study of star formation.  Observational studies of the molecular ISM have shown that it consists of structures over a wide range of sizes, ranging from giant molecular clouds and filaments on large scales (100 pc -- 10 pc) to cores \citep{mol10,and10,jac06} and protostellar disks (0.1 pc -- 0.001 pc) on the small scales.  Fragmentation of the filamentary structures  lead to the formation of dense and cold prestellar cores which eventually evolve into protostars. Due to the range of scales involved and the fact that molecular clouds are highly dynamic in nature, star-formation in molecular clouds is inherently a multi-scale phenomenon with different physical processes (e.g., gravity, turbulence, B-fields, and stellar feedback) having the potential to govern the outcome at different scales. 

While the role of gravity and turbulence have been explored in the past, improved understanding based on polarimetric observations suggest that the B-fields play an important role in the formation of cloud substructure by setting preferred directions for large-scale gas flows in molecular clouds, and also directing the accretion of material onto star-forming filaments and hubs. Subsequent to the formation of a seed protostellar object, accretion as well as outflow are central to the question of assemblage of material from the envelope of the core to form stars. The process of protostellar evolution with the core is relatively well understood for the low mass stars as compared to their massive counterparts \citep{2015IAUGA..2254046T}. In the former, an observational  sequence is clearly established, unlike the latter where multiple factors (including observational limitations) have led to a deficiency in our understanding of the detailed evolutionary sequence. Multiple theories have been proposed to understand the formation of massive stars and it is now largely believed that global collapse and accretion on larger scales can play a major role in their formation, unlike the core-collapse scenario considered for low mass stars \citep{2007ARA&A..45..481Z,2018ARA&A..56...41M}. However, the detailed physical mechanisms which regulate the inward flow of the gas onto the central star, as well as the outward ejection of material in the form of jets and winds from the protostellar system still remain a mystery. The accretion of material onto the protostar through a disk and ejection via jets and outflows are widely thought to be related and episodic events have been observed in both the cases, albeit independently. 

The mass distribution of stars expressed as the stellar initial mass function (IMF) is a fundamental property of star formation, offering key insights into the physics driving the process as well as improving our understanding of stellar populations, their by-products, and their impact on the surrounding medium \citep{2019BAAS...51c.439H}. The physical state of the gas (including temperature, pressure, metallicity, and turbulence) determines which pockets of gas fragment and collapse, and thus ultimately the mass of the stars formed \citep{Hopkins2018}. Based on the stellar IMF, the massive stars (m$_\ast >~$8 M$_\odot$) correspond to only 1\% of the stellar population \citep{massey2003}. However owing to their significant impact on their surroundings  through injection of vast amounts of radiative and mechanical energy in the form of stellar winds, outflows, radiation pressure, photoionization, and supernovae \citep{kennicutt1998}, the feedback from the massive stars govern the star formation efficiency of molecular clouds \citep{elmegreen2011,hopkins2014}. An additional challenge to the star formation studies lies in the fact that stars rarely form in isolation and our Sun, being a single, may be an exception. With nearly half of solar-type stars in the field having a companion \citep[e.g., ][]{2010ApJS..190....1R, 2013ARA&A..51..269D} and the trend being even higher in star-forming regions \citep[e.g., ][]{2016AAS...22720503T}, multiplicity is a ubiquitous feature across the H-R Diagram that deserves significant attention. 

Circumstellar discs or protoplanetary discs surrounding the young stellar objects (YSOs) provide the building material for planet formation. Understanding the physical and chemical nature and the evolution of protoplanetary discs is essential to interpret the planet formation mechanisms and the demographics of the resulting planetary systems. The journey of gas from diffuse atomic through diffuse and dense molecular to protostellar and protoplanetary disks and planets is closely influenced by the astrochemistry occurring in it. Thus, the chemistry of the ISM is intricately related to the initial compositions of planetary bodies, hence are crucial to understand the origin of bio-molecules on planets like the Earth. The molecules detected in the ISM are (i) useful probes of the physical conditions and lifetimes of the environments and (ii) the building blocks of more complex molecules \citep{Herbst2009,Jorgensen2020}. 

This article attempts to capture the cycle of interstellar gas leading to star and planet formation and presents detailed discussions on the following topics:

\begin{enumerate}[label=(\arabic*), nosep]
\item The neutral ISM: From Diffuse Atomic to Giant Molecular clouds
\item Initial conditions for star formation: Filaments to Cores
\item Roles of B-Field \& Turbulence in Star Formation
\item Accretion, Jets \& Outflows
\item Measuring the Low mass end of IMF \& the way forward
\item Feedback of High Mass stars on the ISM
\item Formation and evolution of binary and multiple star systems
\item Planet forming disks and Protoplanets
\item Astrochemistry of ISM, protostars and protoplanets
\end{enumerate} 

The present understanding of the topics of star formation is based on the data from both ground and space-based observatories operating in X-ray, UV, optical, infrared, submillimeter, millimeter and radio telescopes that have also provided a point of comparison for more massive, distant and clustered star formation. The physical conditions in the star forming gas imply a significant role of high-resolution infrared and submillimeter and millimeter observatories, which have become available only during the past decade and a half with observatories such as the Herschel Space Observatory, Atacama Large Millimeter/submillimeter Array (ALMA), GAIA and now the James Webb Space Telescope (JWST). The impact of these observatories is significant in providing an impetus to not only observational studies of star and planet formation but also theoretical and numerical modeling and laboratory astrophysics experiments. 

For each of these nine science topics selected, we describe the status of research at the global as well as the national level, some of the key questions, Indian contribution to the field and the future prospects. Requirement of new facilities or upgrade in the existing ones in terms of telescopes, instruments and computational machines are also discussed.


\section{The neutral ISM: From Diffuse Atomic to Giant Molecular Clouds}

\subsection{Global Status and Open Questions}

The primary mechanism of transformation from warm diffuse phase to cold dense clouds is considered to be the development of non-linear thermal instability triggered by propagation of shocks in the Warm Neutral Medium (WNM; $n \sim 1$\,cm$^{-3}$, $T \simeq 8000$ K) and by converging flow of WNM. Alternatively, a series of shocks and converging flows causes the WNM to break into multi-phased medium with clumpy Cold Neutral Medium (CNM) which is a hundred times denser as well as cooler than the WNM. Other mechanisms for collecting masses,  such as agglomeration (or coagulation) of smaller clouds, converging flows driven by stellar feedback or turbulence, spiral arms passage, cloud-cloud collisions, shock-wave passage and large-scale instabilities have also been proposed. Recent high sensitivity surveys of the Galaxy using multiple tracers of ISM phases and star formation \citep[e.g. THOR survey;][]{beuther2016}, along with advances in theoretical and computational work, have improved our understanding of the ISM and star formation process. Survey of the far-infrared \CII\ emission using {\em Herschel} and Stratospheric Observatory For Infrared Astronomy (SOFIA), as well as comparison of diffuse Gamma ray emission and \HI\ 21\,cm maps have provided evidence for the existence of a component of the dense ISM termed  `dark molecular gas' \citep{wolfire2003} producing little or no CO emission,  possibly comprising of material in transition from diffuse to dense molecular gas. 

There remain many open questions pertaining to the key physical processes in the multi-phase ISM. The ISM is observed to have rich, small scale structures, presumably arising from turbulence \citep[e.g.][]{dickey2001, stanimirovic2010}, however, the details of the nature of turbulence including energy injection/dissipation scales and mechanisms are not well understood. The different theories of formation and evolution of molecular clouds require us to understand the driving mechanisms of supersonic turbulence and how it is sustained. In literature, the proposed mechanisms for supersonic turbulence are B-fields, protostellar outflows, \hii\ regions, supernovae, and on-going mass accretion \citep{Goldbaum2011}.  The first four of these mechanisms have been explored to a certain extent, but the role of accretion of new material in driving turbulent motions is yet to be investigated thoroughly. The cloud mass accretion via atomic gas can be traced using neutral and ionized carbon, which are expected to be the dominant form of carbon in such gas \citep{Heyer2022}. Similarly, in the standard multi-phase model of the ISM, most of the ISM is expected to be in the stable thermal phases within a narrow range of thermal pressure \citep{field1969,wolfire2003}; however the recent observations \citep[e.g.][]{heiles2003,murray2018} of a large fraction of atomic gas in the so-called unstable phase, and its connection with the strength and nature of turbulence in the ISM, are not well understood  \citep[e.g.][]{hennebelle2007,federrath2021}. Recent in-depth study of the filamentary properties of the CNM structures show evidence of  alignment of H~{\sc  i} by small scale magnetohydrodynamic (MHD) dynamo \citep{kalberla20, kalberla21}. These large filamentary molecular clouds of $\sim 100$ pc scales are also found in the Galaxy \citep{Jackson2010}, with some associated closely with nearby spiral arms \citep{Goodman2014}. On the observational front, there are key unsolved questions related to the detailed constituents and properties of the interstellar dust and their connection with the diffuse interstellar band \citep{snow1998}.  There are many useful insights from simulations and observations regarding the formation and evolution of molecular clouds from atomic ISM \citep[e.g.][]{inoue2012,inutsuka2015}, but many important debates are far from being settled. In particular, the relative importance of gravity, turbulence and B-field in this process is yet to be completely understood.

\subsection{Research and Status of the Field in India}

Indian researchers have worked on some of the key questions described above and have made important contributions towards answering them. Several studies aimed towards understanding the thermal phases of the atomic gas using 21 cm observations \citep[e.g.][] {mohan2004,kanekar2011,roy2013,patra2018,basu2022,patra2024} clearly establish the prevalence of atomic gas that is not in stable thermal phase. Studies of the density and the velocity fluctuations to understand the nature of interstellar turbulence, primarily in the diffuse atomic gas \citep[e.g.][]{deshpande2000,dutta2014,koley2019,choudhuri2019}, as well as in supernova remnants \citep{saha2021} have also been performed. Observations and numerical modelling have been carried out to understand the driving mechanism and scale as well as the nature of turbulence on the large scale from the 21 cm observations of atomic hydrogen \citep{nandakumar2023,nandakumar2020,vishwakarma2020}. 
\citet{patra2022} have shown that in the star forming regions located at the outer Galaxy,   the conversion factor between the mass of dense-gas  and line-luminosities of its tracers HCN and HCO$^+$ is quite similar to factors used in the extragalactic studies. Search for potential bones/spurs in the Galactic arms using SEDIGISM CO survey \citep{2017A&A...601A.124S}  has revealed extremely long ($\sim 2-4$\,kpc) scale filament structures like the Gangotri wave \citep{2021ApJ...921L..42V}. Theoretical investigations of the origin and properties of Galactic Cosmic Rays (CR) suggest the possibility that  discrete Supernovae explosions may be responsible for CR acceleration and  find that an injection rate of one supernova in every $\sim 0.03$\, Myr can explain the $\gamma$-ray profile observed \citep[e.g.][]{bhadra2022}. Direct measurements of the ISM B-field strength have also been done using tracers of different density ranges \citep{koley2021,koley2022}. 


\subsection{Key Science Questions and Future Goals}

The relative distribution of gas in the different phases of the ISM \citep{roy2017}, and whether the ratio of the gas in different phases is constant or evolving has been a long standing question. There is mounting observational evidence for previously undetected phases of gas in the form of as the CO-dark diffuse molecular gas or in H~{\sc i} self absorption, however understanding of the dependence of physical conditions is still an open issue. A rather long term effort would be to trace out the detailed morphological and dynamical structure of the Galaxy, including different atomic and molecular phases. This would include the organisation and characteristics of H~{\sc i} and molecular filaments, the inter-cloud collisions and their effect on star formation, formation and evolution of molecular clouds, characteristics of turbulence and its relation to dust and B-field structures. While turbulence is understood to play a major role in determining the dynamics and morphological evolution of the Galaxy, the relative importance of the different driving mechanisms over different length scales and their interplay require investigation. These need to be complemented with studies of the B-fields in diffuse atomic and molecular phases, possibly using large scale Zeeman surveys with H~{\sc i}, CH, CS, CCS along with polarised continuum (DCF) to trace the 3D B-field structure. 

Addressing these question requires both theoretical investigations and numerical simulations incorporating different physics as well as well-planned observational investigations using multiple tracers \citep[][]{lauroesch1999}  along the same or nearby lines of sight \citep[][]{roy2006}. Along with the targeted studies, long term projects like Low frequency survey of the Galactic plane using the uGMRT,  Dispersion Measure mapping using the uGMRT can be considered as important aspects of addressing new discoveries.


\section{Initial conditions for star formation: Filaments to  Cores}

\begin{figure*}
    \centering
    \includegraphics[width=0.95\textwidth]{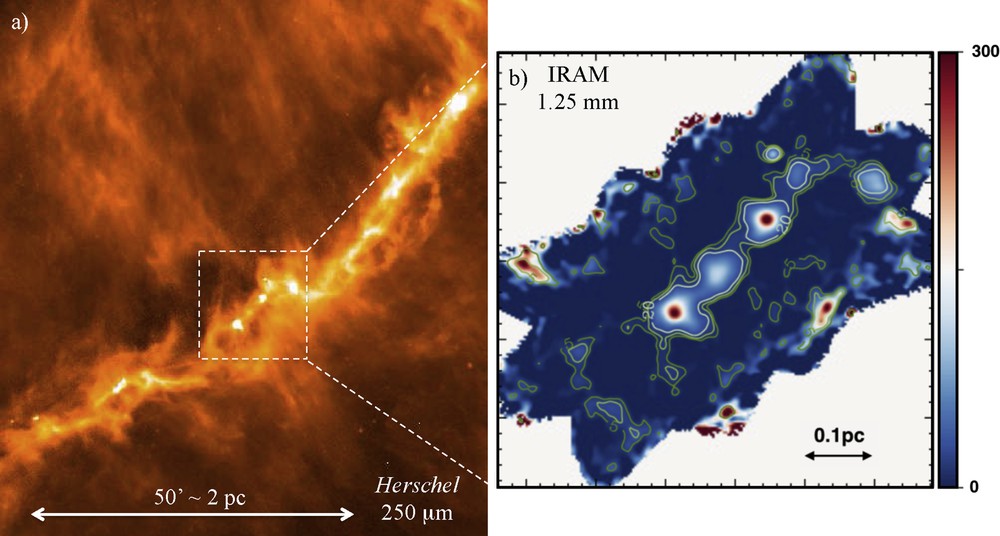}
    \caption{(a) Herschel/SPIRE 250\,\micron\ dust continuum image of the Taurus filament, b) IRAM 1.25-mm dust continuum image, showing star forming cores (adopted from \citealp{and17}).}
    \label{fig_fc_01}
\end{figure*}
\begin{figure}
	\centering
	\includegraphics[width=0.45\textwidth]{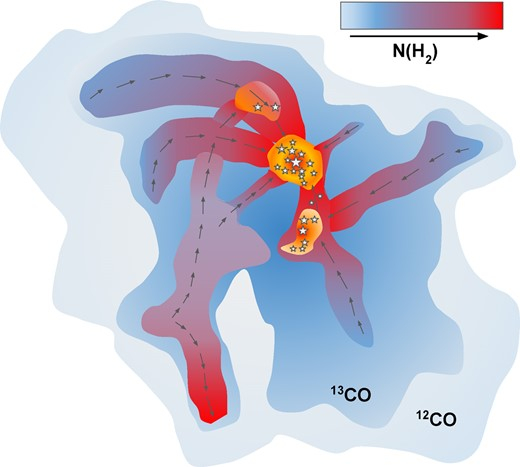}
	\caption{Cartoon illustrating formation of star clusters at the nexus of filamentary flows in the Galactic giant molecular cloud  G148.24+00.41 (adopted from \citealt{rawat24}).}
	\label{fig_fc_02}
\end{figure}

Several surveys of the Galactic plane and nearby star forming regions have revealed the near ubiquitous presence of filamentary molecular clouds. The fragmentation of these filamentary structures in the cold molecular ISM leads to the formation of dense and cold prestellar cores (shown in Figure~\ref{fig_fc_01}) which eventually evolve into protostars. Further, some filaments intersect at locations called ``hubs'' (Figure~\ref{fig_fc_02}), which are the preferred sites for high-mass star formation due to high surface density on account of mass being channeled towards hubs through the filaments. One overarching task for understanding star formation in hubs and filaments is to understand how the different scales are connected and which physical processes dominate on which scales, and to establish a universal route to star formation from giant molecular clouds. 

\subsection{Global status, and open questions}
Understanding the initial phases of star formation requires characterization of the coldest and densest parts of molecular clouds (e.g., density and temperature profile, kinematics, chemistry, etc.). In the last decade, several dedicated wide-field surveys either in dust continuum (primarily in the 800-1100\,\micron\ range) or molecular line emission (mainly CO isotopologues) sampling the Galactic plane and a number of nearby clouds have been carried out using single-dish facilities such as ATLASGAL and SEDIGISM surveys with 14-m APEX (Chile); BGPS with 10.4-m CSO (Hawaii); JPS, GBS, COHRS, CHIMPS with 15-m JCMT (Hawaii); GRS with 14-m FCRAO (Massachusetts); FUGIN with 45-m Nobeyama (Japan); MOPRA-CO and ThrUMMS with 22-m MOPRA (Australia), and MWISP with 14-m PMO (China).  Details of all the aforementioned surveys can be found in \citet{sch21}. Similarly, several targeted observations or mini surveys have been taken up using high-resolution interferometric facilities such as with ALMA in Chile, NOEMA in the France, and SMA in Hawaii \cite[e.g.][]{mot22}. All these high cadence surveys along with the results from Herschel observations have significantly improved our understanding of star formation. A few notable results from these surveys are: (i) only a small fraction of the GMC material ($<$ 20\%) lies in filaments and participate in star-formation, (ii) core-to-star formation efficiency is likely between 20\% to 40\% \cite[e.g.][]{and10,kon15} (iii) the origin of the IMF in the range $\sim$0.1 to $\sim$5 M$_\odot$ is likely driven by turbulence \citep{and17}, (iv) the growth of clumps/cores that are embedded within supercritical filaments is related to the gas dynamics, both perpendicular to and along the filament \cite[e.g.][]{per14}, and (v) filaments themselves are composed of bundles of closely spaced velocity coherent structures, called fibres with distinct physical properties and low-mass cores are tied with these fibres \citep{hac18}. 

On the theoretical side, multi-scale numerical simulations, incorporating some of the physical processes for a range of reasonable initial conditions with and without B-fields have also been carried out world-wide and have been extremely helpful in explaining the formation of different types of filaments and the net outcome of star formation with them  \citep[e.g. see][and references therein]{pin22}. 

Despite the significant progress, there are a number of questions, for which no global consensus has yet been  reached. These include, but are not limited to: (i) What controls the fragmentation properties (thermal vs. turbulence) of super-critical filaments? (ii) Are the fragmentation properties of filaments scale dependent? (iii) How do massive supra-jeans cores form inside the filaments for massive star formation to occur?  (iv) Are dense cores isolated objects as they are located within hierarchically small-scale structures such as fibres and streamers? (v) What are the density profiles of the pre-stellar cores, and how do they evolve with time? (vi) What is the life-time of pre-stellar cores, and how does the chemical evolution of cores proceed? (vii) Do isolated cores and cores in massive clumps have the same profile, lifetime, and chemical clock?

\subsection{Research and status of the field in India}
The star formation processes can only be studied using sub-mm observations which can probe such cold, dense environments. Due to the lack of sub- and millimeter band facilities in the country, over the last decade, the involvement of Indians in this field is mostly based on available low-resolution archival data sets, focusing primarily on the structures and kinematics of the large-scale structures of the ISM and star-formation within them. These studies are important in their own way and have improved our understanding  of a number of key questions such as the formation of dense gas and star formation within them at the nexus of hub filamentary system  \citep[e.g.][]{bag18,dew21,rawat24} or  cloud-cloud collision \citep[e.g.][]{dew18,iss20,dey22}, influence of bipolar \hii\ regions on the star-formation processes in the ambient filamentary molecular clouds \citep{sam18,mookerjea2022}, likely large-scale collapse processes (e.g. edge collapse) of the filaments \citep[e.g.][]{dut18,dew19a},  star formation processes in the inter-winded filaments \citep{dew21b}, role of B-fields compared to gravity and turbulence on the stability of dense structures in the filamentary environment \citep[e.g.][]{esw20,eswar2021,sha20,sha22}, and likely fragmentation processes of the filaments \citep[e.g.][]{sam15,veena18}.  Recently, a few studies have probed star formation at the level of clumps and cores using high spatial resolution observations with ALMA and SMA.
These include the physio-chemical properties of low-mass cores \citep[e.g.][]{sahu18}, accretion process in the vicinity of  massive young stellar objects \citep[e.g.][]{dew22}, and role of turbulence on the star-formation processes of the cores \citep [][]{bag21,saha22,pandian24}. \citet{Dewangan2024g11} used JWST NIR images to discover an IR-dark HFS candidate (extent $\sim 0.55$\,pc) towards the previously known massive protostar G11P1.

\subsection{Key science questions, future goals etc from an Indian perspective.}

A few key questions pertaining to the formation of star or star-cluster in the filamentary structure of the ISM include: (i) how exactly does the  accretion occur in the filamentary environment from large-scale to the smallest scale? (ii) how to refine the fraction of matter in a GMC that resides in filaments, clumps and cores?  (iii) how to derive better constraints on the evolutionary status and evolutionary history of clumps/cores by tracing their chemical composition? (iv) what is the nature of turbulence at different scales and its role in making structures and star formation within them compared to B-field and gravity? and (v) how to test various modes of cluster formation, through kinematics and demographics of cores in cluster forming clumps? The answers to many of these questions lie in the understanding of how the physical, kinematic, and chemical properties are related and how they vary at different scales for a wide variety of environments (e.g. the Central Molecular Zone, the outer Galaxy), and how the relationships between different scales influence the outcome of star formation processes. These goals necessitate observation of a large-number of molecular clouds using both single-dish as well as interferometric sub-mm/mm facilities in multiple molecular lines and thermal dust continuum. Indian astronomers are working on a number of puzzling questions that exist in the field, but in pieces. A dedicated survey aiming to map a few molecular clouds in a coherent way from large-scale to small-scale would be highly valuable in this regard and would be able to provide information on both physical and dynamical properties of the gas, tracing gas motions and angular momentum from cloud to core scale.

\section{Role of Magnetic fields and turbulence in star-formation}

\subsection{Global Status and Open Questions}
B-field and turbulence are found to play crucial and interesting roles in star formation and evolution of molecular clouds and cores where stars form \citep{pattle2022}. However a complete understanding of the individual and relative importance of these two components vis-a-vis gravity is still rather incomplete. In the recent years there has been a significant increase in the number of efforts to understand the B-field and turbulence in the ISM.  B-fields are found pervading all different phases of the Galaxy \citep{fer2020}. Numerical simulations \citep{SF2022, GentEA2023} and observations \citep{BraccoEA2020, BorlaffEA2021} show that properties of B-fields differ with ISM phases. A number of observational studies are now available where B-fields are mapped and their properties are studied in different star forming regions at different evolutionary stages and in different environments. 
\citet{soam2024} have recently reviewed observations, numerical models and relevant instrumentation that address the role of B-fields and turbulence in the formation of stars.

\begin{figure*} 
\centering
\includegraphics[width=0.7\textwidth]{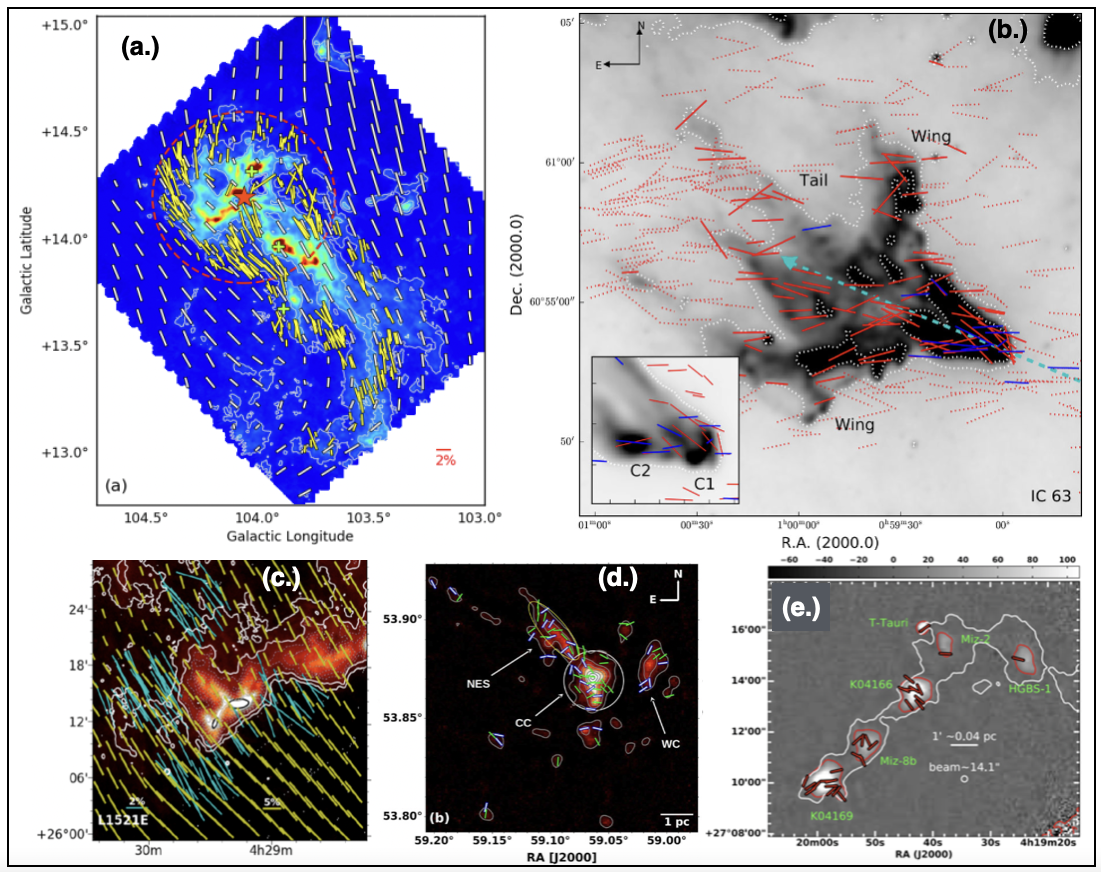}
\caption{A compilations of B-field morphologies traced in different regions of the ISM. These regions include (a.) bright-rimmed cloud (BRC) L1172/1174 cloud complex \citep{saha2021}(b.) BRC IC 63 in the vicinity $\gamma$ Cas \citep{soam2017}, (c.) Starless core L1521E \citep{sharma2022}, and (d.) Giant Molecular Cloud G148.24+00.41 \citep{rawat2024} and, (e.) Taurus cores \citep{eswar2021}.}\label{fig:MF_all}
\end{figure*}

Similar to B-fields, roles played by turbulence in star-formation is found to be a crucial thing to explore. Turbulence is found to play dual role, on the one hand it counterbalances gravity and prevents collapse of cores to form stars, on the other in the cold and dense ISM, supersonic turbulence leads to shocked gas compression facilitating star formation. \citet{larson1981} reported a positive correlation between size scale and velocity dispersion in the sites of star-formation. There are a number of studies supporting this correlation \citep{maclow2004, krumholz2005, elmegreen2004, hennebelle2008}. The creation of turbulence can be a result of mechanisms such as accretion, spiral shocks, galactic shear, magneto-rotational instability, and more generally gravity-driven turbulence \citep{KEA2018}.

Several numerical models have been developed to understand the role of B-fields and turbulence in star formation \citep{BrucyEA2020,GrudicEA2021,MFS2022}. Some of these models explore the primary reasons behind inefficient star formation and the dependence of IMF and rate of star formation on the mode of turbulence \citep{federrath2021,MFS2022}. One of these models on star cluster formation has reported that the inclusion of outflows may affect the results by reducing the star-formation rate, enhancing fragmentation and affecting IMF to the lower mass side \citep{MF2021}.

Polarization of starlight and polarized thermal dust emission are found to be the powerful tools to map interstellar B-fields in different wavelengths. In last many years, various observing facilities secured polarimeters for measuring polarization signals on different scales of star forming regions. These instruments covers polarization measurements from optical to sub(mm) wavelengths. Some of these instruments which were used to observe polarized signal towards several regions are ARIES Imaging Polarimeter (AIMPOL) on the Sampurnanand telescope \citep{rautela2004}, SIRPOL on the InfraRed Survey Facility (IRSF; \citet{kandori2006}), James Clerk Maxwell Telescope (JCMT)’s POL-2 polarimeter \citep{friberg2016} on the SCUBA-2 camera \citep{holland2013} operating at 850\,$\mu$m and 450\,$\mu$m, Balloon-borne Large-Aperture Submillimetre Telescope for Polarimetry (BLASTPol: 250\,$\mu$m, 350\,$\mu$m, 500\,$\mu$m \citep{galitzki2014}, Mimir on the Perkins Telescope \citep{clemens2007}, SOFIA's HAWC+ camera \citep{harper2018} operated in five bands from 53\,$\mu$m to 214\,$\mu$m, and ALMA \citep{cortes2016}.

\subsection{Status of research in India}

The Indian community has contributed significantly to understanding B-fields in the ISM. We used the limited available facilities, such as the AIMPOL and the IUCAA Imaging POLarimeter (IMPOL), to produce B-field maps and investigate their properties in different regions of the ISM \citep[e.g., ][]{soam2019, eswar2019, eswar2020, bijas2023, sid2023}. See Figure \ref{fig:MF_all} for some of the results compiled from those and other studies. Later, the community extended the work towards understanding grain alignment theories using the observed results \citep{Soam2021b}. We are developing more instruments like AIMPOL and IMPOL, including NICSPol \citep{arthy2019} and EMPOL \citep{ganesh2020} at Mt. Abu observatory. Indian researchers are involved in larger projects worldwide where attempts are made to map B-fields in the Galactic plane using instruments like JCMT/POL-2. More details on available, ongoing, and upcoming polarimeters in India are included in a recent study by \citet{soam2024}.

\subsection{Key Science Questions and Goals}
The key science questions in star-formation is to pin-point the exact roles of B-fields and turbulence in this process at different scales and in different environment. There is a lot of progress on mapping B-fields in the ISM and investigating the energy budget (i.e. comparison between magnetic, turbulent, and gravitational forces) in those regions. Several large surveys across the world are ongoing these days to study B-field properties using polarization in multi-bands. Multi-wavelength observations help in tracing field properties at different spatial scales. Some important areas however require further investigation \citep[cf references in][]{pattle2022, soam2024}.

These include answering some questions on how B-fields and gas couple together \citep{mestel1966} and govern various physical processes in the ISM. How B-fields actually help in cloud formation and fragmentation of long filaments into smaller dense cores where stars form? How are the circumstellar disks formed and how are the outflows launched? Does B-field have a role in these phenomena? Is it B-fields alone or are there other agents which interact with B-fields and aid star-formation? 

\section{Accretion, Jets and Outflows}

One of the first signatures of the early phases of protostellar formation are jets and outflows which are easy to locate based on the effects that they cause in the surrounding parent cocoon. They are observed across the entire stellar mass spectrum, from brown dwarfs to massive stars suggesting a universal launching mechanism \citep{2021NewAR..9301615R}. They are believed to carry away and dispel excess angular momentum from the protostellar system
to the outer regions. There have been numerous surveys carried out to characterise jets, and outflows \citep{2016ARA&A..54..491B}. Most of these are carried out using kinematics of molecular line tracers. Jets have mostly been detected through the shocks they produce when they interact with the ISM, and have been often used as a pointer for active star formation in a cloud.

It is widely accepted that the jets are associated with magnetised disks \citep{2004ApJ...612..342A}. This is because disks, unlike jets and outflows, are more difficult to observe directly as they require very high angular resolution imaging. Although disks are known to exist for a substantial part of the protostellar and early main-sequence phase, here we will focus only on
the early phases where active accretion of envelope material is occurring in the disks. The accretion process occurring in young protostellar systems are often probed through strong emission lines of hydrogen and calcium \citep{2016ARA&A..54..135H}. The accretion of material onto the protostar through a disk and ejection via jets and outflows are widely thought to be
related and episodic.

\subsection{Global Status and Open questions}

The accumulation of envelope material to protostars and the ejection in the vicinity through jets and winds are being actively pursued world-wide through indirect observational measures and simulations. This is because it is relatively difficult to probe in the  immediate vicinity of the protostars on the scales at which these operate, which is about tenths to tens of AUs \citep{2014prpl.conf..101L}. The inflow of gas from the envelope to a disk-like structure has been investigated and few studies have claimed to have detected the centrifugal barrier that marks the flow of gas from the envelope to the disk around the protostar \citep{2014ApJ...791L..38S,2018A&A...617A..89C}. Efforts are on to build and carefully characterise the accretion, disk mass and protostellar properties of large samples of YSOs in Class 0 and Class I phase \citep{2022arXiv221107653F}. Accretion rates of material flowing into the protostellar photosphere, and imprinted in spectra are being used to infer the properties of the infalling gas \citep{2022arXiv221106454C}. FUors and EXors are classes of objects that are being investigated to understand the accretion related outburst phenomena in low-mass pre-main sequence stars \citep{2014prpl.conf..387A}.

Some nearby low-mass disk-jet systems have been examined at high angular resolution using ALMA, and diverse properties have been determined such as super-fast rotation ($> 20$~km/s) of jets \citep{2021ApJ...916...23M}, toroidal B-field using molecular line polarisation measurements \citep{2018NatCo...9.4636L},  radial flow close to launch suggesting a magnetic-centrifugal launch of jets \citep{2022ApJ...927L..27L}. In jets from massive protostars, ionisation fraction has been ascertained which helps understand the coupling of matter with  B-field \citep{2019NatCo..10.3630F}. Ionised gas measurements are also being investigated in massive jets near the collimation zone indicating the presence of a wide-angled wind with a collimated jet \citep{2022ApJ...931L..26R}. However, these are isolated examples and a larger sample is necessary to gain a comprehensive understanding of these processes. Regarding the assembling of matter in the protostellar stage, studies at multiple scales are beginning to indicate that for massive star-formation, the central high-mass protostar, the core, and the clump accrete matter and grow in mass simultaneously \citep{2018ApJ...852...12Y}. Studies of simultaneous infall and outflows from massive cores suggest gravitational instabilities, fragmentation, and episodic accretion events associated with disks \citep{2022A&A...659A..81B}. There is a large volume of effort along these directions to understand the earliest phases of star formation, particularly for massive stars.

Some of the open questions in the topic are: How does accretion proceed from envelope to disk and disk to protostar? In other words, how does the protostar accumulate mass from the envelope? How does the protostellar mass and luminosity evolve with time? How is the angular momentum of the protostellar system transported out? Is it via winds/jets alone or through B-field as well? What decides the relationship between accretion and outflow? Both the accretion and ejection of material are found to be episodic. What prompts this? What is the exact launching mechanism (magnetic-centrifugal or magnetic pressure-driven) of jets? At what distance from the exciting source does the wind / jet get initiated? What factors are responsible for the collimation of the jet? Is it related to inherent launch or it is related to pressures from the surrounding medium farther from the exciting source? Rotation and B-fields are central to the problem of accretion and outflow of material. What decides the relative alignment of these axes and the effects on the accretion and outflow?

\begin{figure*}
	\begin{center}
		\includegraphics[width=7cm]{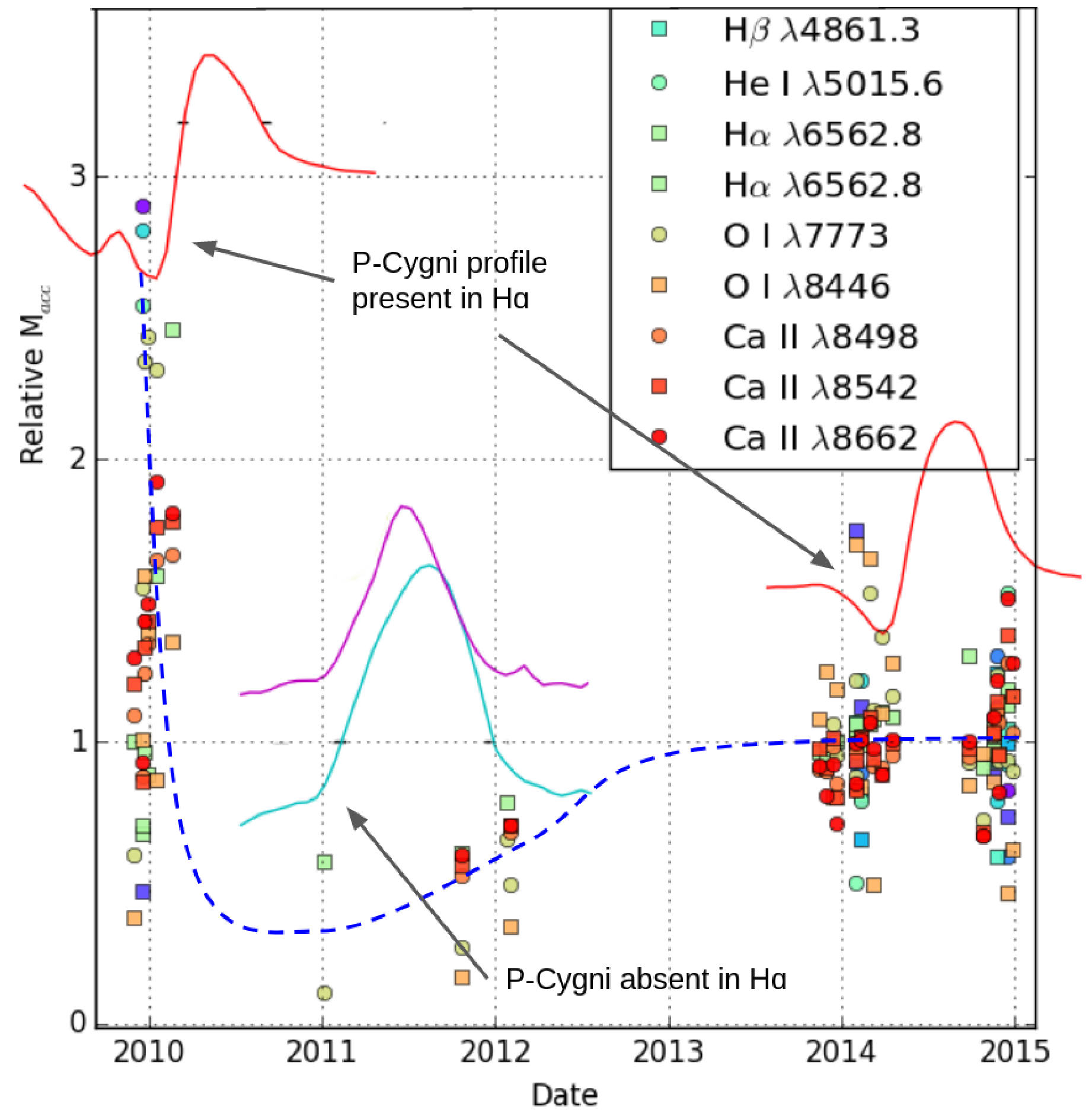}
		\includegraphics[width=8cm]{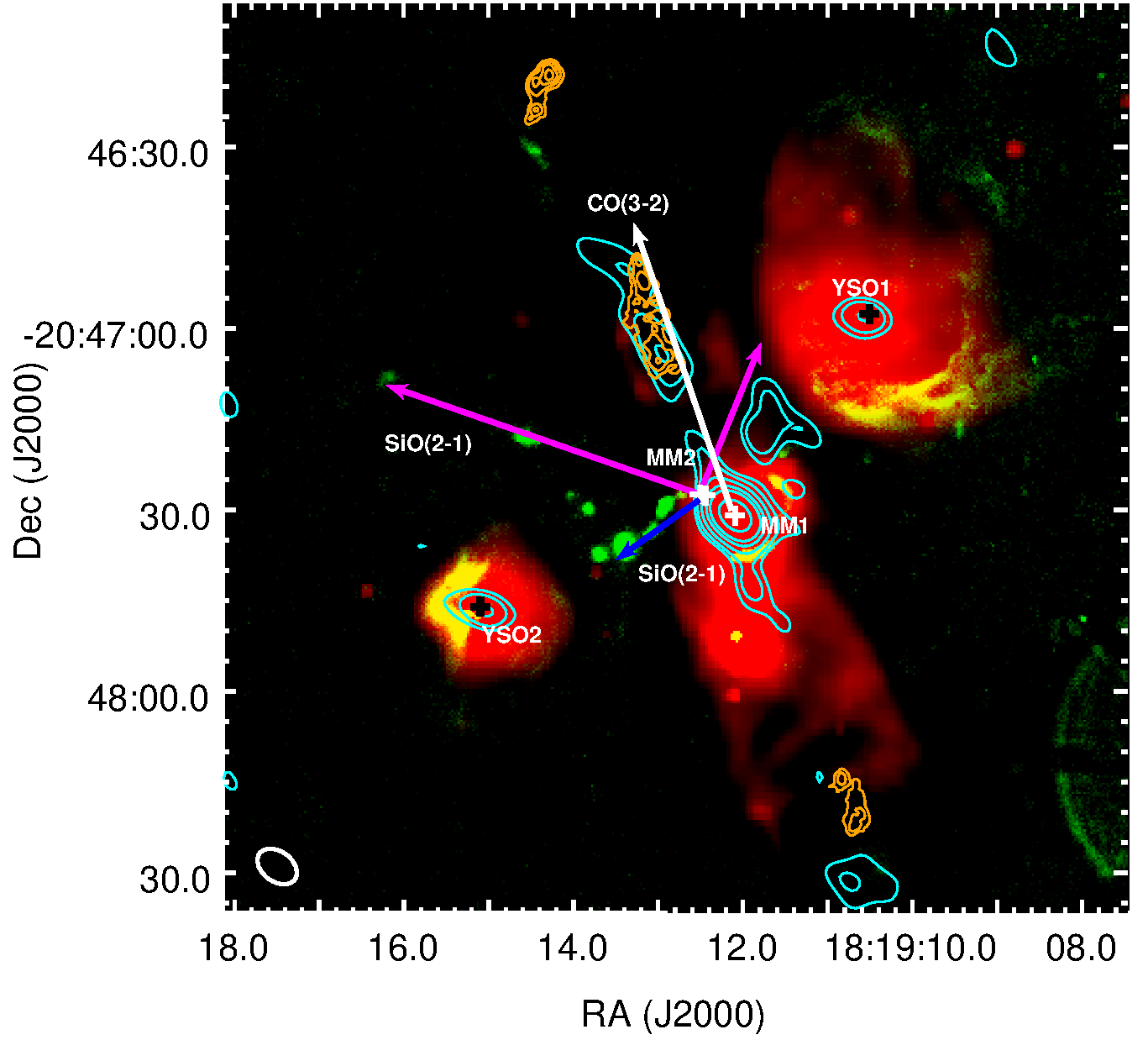}
		\caption{(Left) Discovery of wind signature (traced by the P-Cygni profiles) disappearing during the short period of quiescence in an episodically accreting protoplanetary disc	system (V899 Mon). The wind profile reappeared after the target came back to its high accretion rate \citep{2015ApJ...815....4N}. (Right) Two color-composite image of the central region of the massive protostellar jet HH80-81 jet. The Spitzer 8~$\mu$m broadband image is shown in red and the NIR H$_2$ emission is shown in green. The NIR [FeII] emission is overlaid as orange contours, cyan contours show ionized gas at 1300 MHz, and the arrows depict outflows detected in this region. The white crosses represent massive cores that are believed to drive some of the outflows \citep{2023ApJ...942...76M}.
		}
		\label{fig:accre_jet}
	\end{center}
\end{figure*}

\subsection{Research and Status of the field in India}

There are multiple groups working in various areas of star-formation in India. 
Accretion around low-mass stars have been investigated through variability of emission lines indicating episodic accretion in disks around low-mass pre-main sequence stars and FUors. India has a multi wavelength long-term monitoring program of FUors and EXors (MFES). These are protoplanetary disc objects that are detected to be undergoing sudden episodic increase in its accretion rate by a factor of $\sim100$. This program has resulted in many discoveries, like the first observational evidence that the heavy winds near the magnetosphere are powered by the accretion, see Figure~\ref{fig:accre_jet} (left) \citep{2015ApJ...815....4N}. This program also showed the conventional disc instability models cannot explain the short breaks seen in the outbursts of V1647 Ori, and V899 Mon \citep{2013ApJ...778..116N}, hinting at some  other unknown instability in the magnetosphere for causing the disruptions in the high rate of accretion. New members of this variable class of FUors/EXors also were discovered and characterised under this program \citep{2022ApJ...926...68G}. These discoveries were all obtained using an array of Indian observational facilities like HFOSC, NIRSPEC and TIRSPEC on 2-m Himalayan Chandra Telescope, IFOSC on 2-m Girawali Optical Telescope, and  TANSPEC and TIRCAM2 on 3.6-m Devasthal Optical Telescope.
The accretion around protostars using CO as tracers have also been studied as a function of age \citep{2016ApJ...831...69M}.  Massive protostars, on the other hand, are being investigated through the characterisation of their outflows and possible disks using the ALMA; see \citet{2020ApJ...890...44B}, \citet{2022ApJ...925...41D}, and references therein for some representative work. In addition, high-resolution near-infrared imaging has been used to characterise jets \citep[eg. ][]{2015ApJ...803..100D,2023ApJ...942...76M}, see Figure~\ref{fig:accre_jet} (right) for an example. The GMRT is proving useful in examining ionised gas from protostellar jets at low frequencies. Both thermal \citep{2019MNRAS.485.1775I} and non-thermal emission \citep{2018MNRAS.474.3808V} from jets have been detected using the GMRT. For these ionised jets, a numerical model has been developed by \citet{2022MNRAS.514.3709M} to characterise the physical properties of the jet by constraining the radio spectral indices. The footprint of imaging studies involving accretion phenomena and jet-launch close to the protostars has been relatively small in India, largely due to the lack of access to high resolution (sub-arcsec resolution) imaging telescopes, particularly at sub-millimetre wavelengths. 

\subsection{Key science questions and future goals from an Indian perspective}

The key science questions related to assemblage of matter onto a protostar are challenging even from a global perspective. As mentioned, early results from facilities capable of high resolution observations have just started to come in. Based on available facilities, the Indian perspective would be to understand the infall and outflow rates by characterising the accretion as well as jets and outflows from protostars. This is because it is relatively easier to study accretion through the kinematics of emission lines, and jets / outflows via their kinematics as well as interaction with the ambient interstellar medium. Multi-wavelength studies will play a crucial role here and studies in optical, infrared, radio and X-rays can be combined to give a coherent picture of the phenomena under study. The studies in the past have largely been based on isolated cases, and there is a need of the hour to have large surveys with collaborative efforts in order to get a global picture of these processes. A comparison between the accretion and outflow rates in large number of young protostellar systems can shed light on the mass census of the protostars. In addition, the evolution of the protostellar systems through the outflow properties can also be ascertained.

\section{Measuring the lower mass end of the IMF \& the way forward}

The physical state of the gas (including temperature, pressure, metallicity, and turbulence) determines which pockets of gas fragment and collapse, and thus ultimately the mass of the stars formed\footnote{https://ned.ipac.caltech.edu/level5/March18/Hopkins/Hopkins2.html}. Thus, the stellar IMF is a fundamental property of star formation, offering key insight into the physics driving the process as well as informing our understanding of stellar populations, their by-products, and their impact on the surrounding medium \citep{2019BAAS...51c.439H}.

\subsection{Current status and open questions}
The main focus in the study of the IMF is on constraining two of the most important characteristics of the IMF: (a) the shape of the IMF at lower mass end and (b) the form of the IMF in different local conditions. The theory has long predicted a significant boundary of star formation around 1 to 10 Jupiter mass, below which it is believed that fragmentation and collapse of molecular cloud cores cannot occur. However, observations show that the IMF is almost feature-free, all the way from the most massive stars down to 0.3-0.5 M$_\odot$. Below 0.3-0.5 M$_\odot$, studies have shown that the slope of the  IMF becomes shallow and below 0.1\,M$_\odot$ (100 Jupiter mass), it starts to decline. Hence, whatever be the nature of physical processes governing star formation, they produce this turnover and decline of the IMF, with a likely boundary condition at some lower limit. This further leads to the implication that at the low mass end, the current fragmentation paradigm might need to be replaced by a model which incorporates diverse physical processes, such as supersonic turbulence \citep{2002ApJ...576..870P}, feedback effect of bipolar outflows \citep{1996ApJ...464..256A} and ionizing radiation from massive stars \citep{2000ApJ...540..255P}, and dynamical interactions between protostars \citep{2002MNRAS.336..705B}, to mention a few. Subsequently, it is reasonable to conclude that the form of the substellar IMF governed by these processes could further our understanding of the star formation process itself \citep{Gardner2006}.

\begin{figure*}
    \centering
    \includegraphics{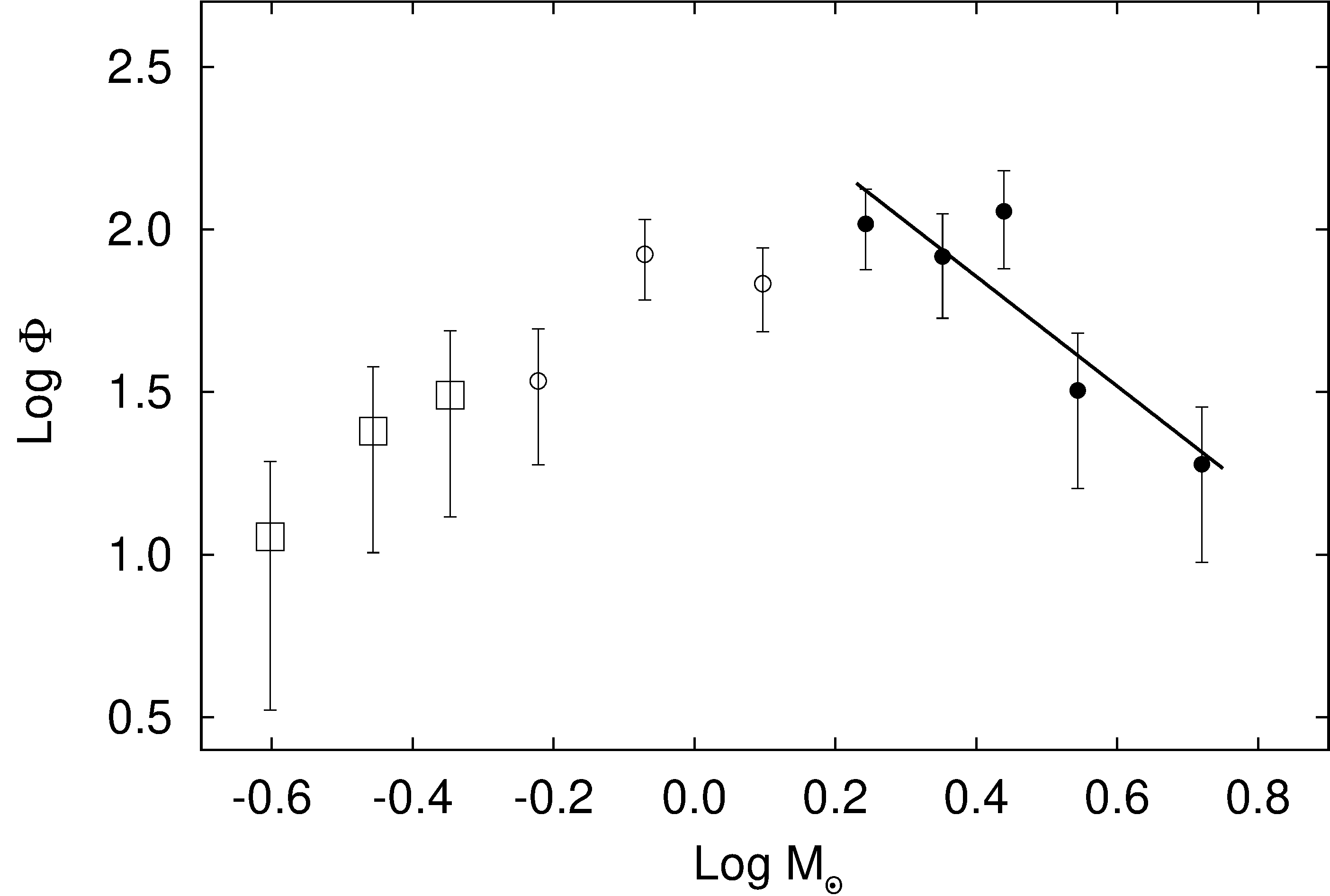}
    \caption{A plot of the MF for the stellar sources in the Mayer 3 cluster region.
Log $\phi$ represents log($N$/dlog $m$). The error bars represent $\pm\sqrt N$ errors. The solid line shows the least squares fit to the MF distribution (black dots).
Open squares are the data points falling below the completeness limit of 0.8 M$_\odot$. Open circles are the data points near turn-off point in the MF distribution and are not used in the fitting. For more details, please refer to the original publication by \citet{pandey2020}. This figure is reproduced by permission of the AAS. For more details see \citet{pandey2020}.}
    \label{fig:imf}
\end{figure*}

Numerous studies have given hints of the existence of cluster-to-cluster variations in the IMF across the entire stellar mass range. As no statistical study covers different homogeneous local conditions, it is not yet known how the IMF might behave across a wide range of environments, such as those with extreme gas temperatures and densities, high pressures, and low metallicities.  

\subsection{Research and status of the field in India}
 
Numerous studies on mass function, based on the sample of young and old open star clusters have been already done, but for mostly higher mass bins ($>$1 M$_\odot$). Recently, using wide-field deep optical \& NIR data of young star forming regions, the IMF slopes have been constrained upto $\sim$0.5 M$_\odot$ \citep[cf.,][and references therein]{2008AJ....135.1934S,2009ApJ...693..634O,2012PASJ...64..107S,2014MNRAS.443.3218M,2017MNRAS.467.2943S,2017ApJ...836...98J,2012MNRAS.424.2486J,2020MNRAS.498.2309S,2020ApJ...896...29K,pandey2020,Damian2021,2022JApA...43....7P}. 
The main results of these studies are: (a) there is a change in mass function slope at 1--1.5 M$_\odot$ also, (b) at higher mass bins (1 - 10 M$_\odot$), there exists a variation of mass function slope in different regions of our Galaxy (d $<$ 3 kpc) and (c) massive stars are generally formed in the inner part of the clusters. Figure\,\ref{fig:imf} shows the MF observed in Mayer 3 star cluster where the MF slope is found to be $\Gamma\simeq$-1.7, steeper than the Salpeter value of  $\Gamma$=-1.35 \citep{1955ApJ...121..161S} in  a mass range of $1.5<M_\odot<6.5$, indicating the abundance of low-mass stars,  probably formed due to the positive feedback of the massive stars in this region \citep{pandey2020}. There is also a change of MF slope from the high to low mass end  with a turn-off at around 1.5 M$_\odot$. A truncation of MF slope at slightly higher-mass bins has often been noticed in other SFRs also under the influence of massive OB-type stars \citep{2007MNRAS.380.1141S, 2008MNRAS.383.1241P, 2008MNRAS.384.1675J,2017MNRAS.467.2943S}. In a recent analysis \citet{Damian2023} obtained the sub-stellar IMF of Sigma Ori cluster down to Jupiter mass regime. 

Since most of these studies were carried out using a meter class telescope on a sample of nearby star clusters (d$~<3$ kpc), the subsolar limits of IMF were not yet reached and lack statistically significant sample set representing different local conditions.  Also, these studies were done on a large variety of data sets, thus the homogeneity of the data set is a big question on conclusions regarding the universality of the IMF slopes.

%

\section{Impact of massive stars on the interstellar medium}
 Stellar feedback has profound impact on the star formation efficiency of molecular clouds \citep{elmegreen2011,hopkins2014}. Feedback from a massive star can either shred the nascent molecular cloud within a few cloud freefall time-scale and halt further star formation \citep{kim2018} or can trigger formation of the next generation of stars by creating physical conditions (e.g. high densities, instabilities due to shockfronts) that leads to the collapse of a molecular cloud that otherwise may not contract and fragment spontaneously \citep{elmegreen1977}. The extreme ultra-violet (UV) photons  from the massive stars ionize the surrounding medium and create
\hii\ regions, while the far-ultraviolet (6$<$h$\nu <$13.6\,eV) photons create the Photo Dissociation Regions (PDR), a layer of warm atomic and molecular gas between the \hii\ region and the molecular environment \citep{hollenbach1999}.
Additionally, the massive stars through their winds and supernova explosions tremendously impact the dynamical evolution of ISM on all scales by generating turbulence starting from sub-parsec scale (radiation pressure and jets and outflows) to Galaxy scale (supernovae explosions). It is also believed that the latter is possibly responsible for maintaining gas at large scale heights that resist gravitational collapse onto the host galaxy's potential \citep{mckee1977}.

\begin{figure*}[!t]
\centering
\includegraphics[width=5.2cm]{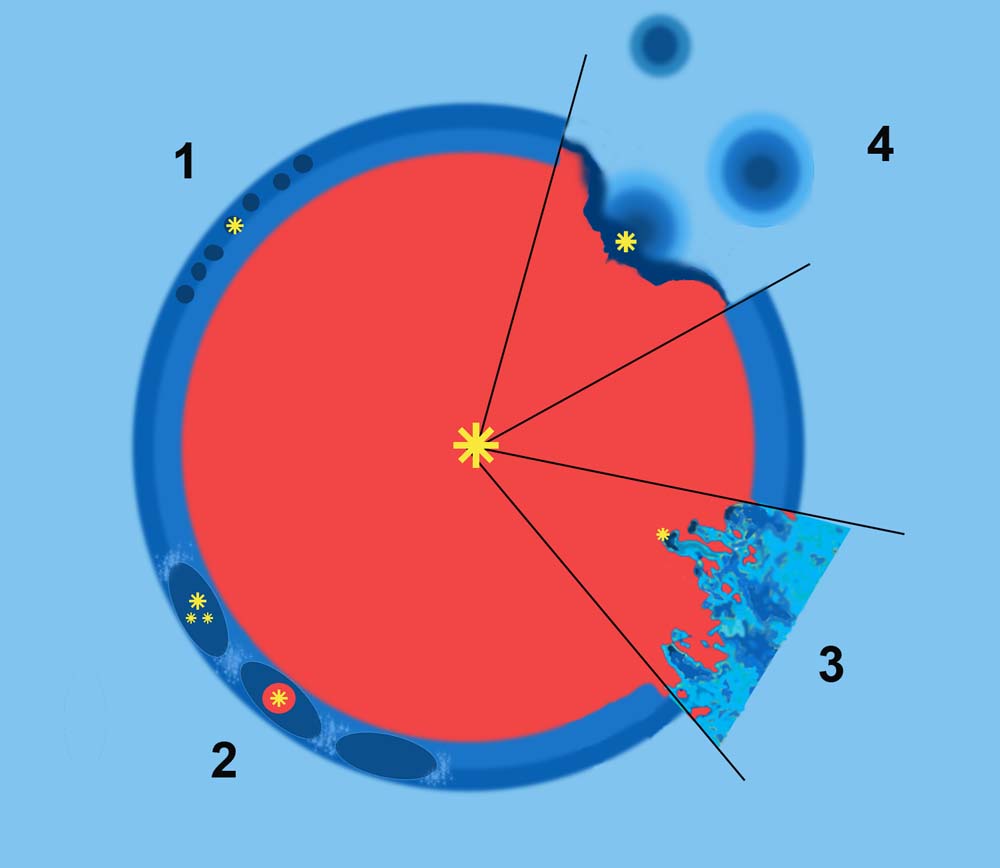}
\includegraphics[width=5.7cm]{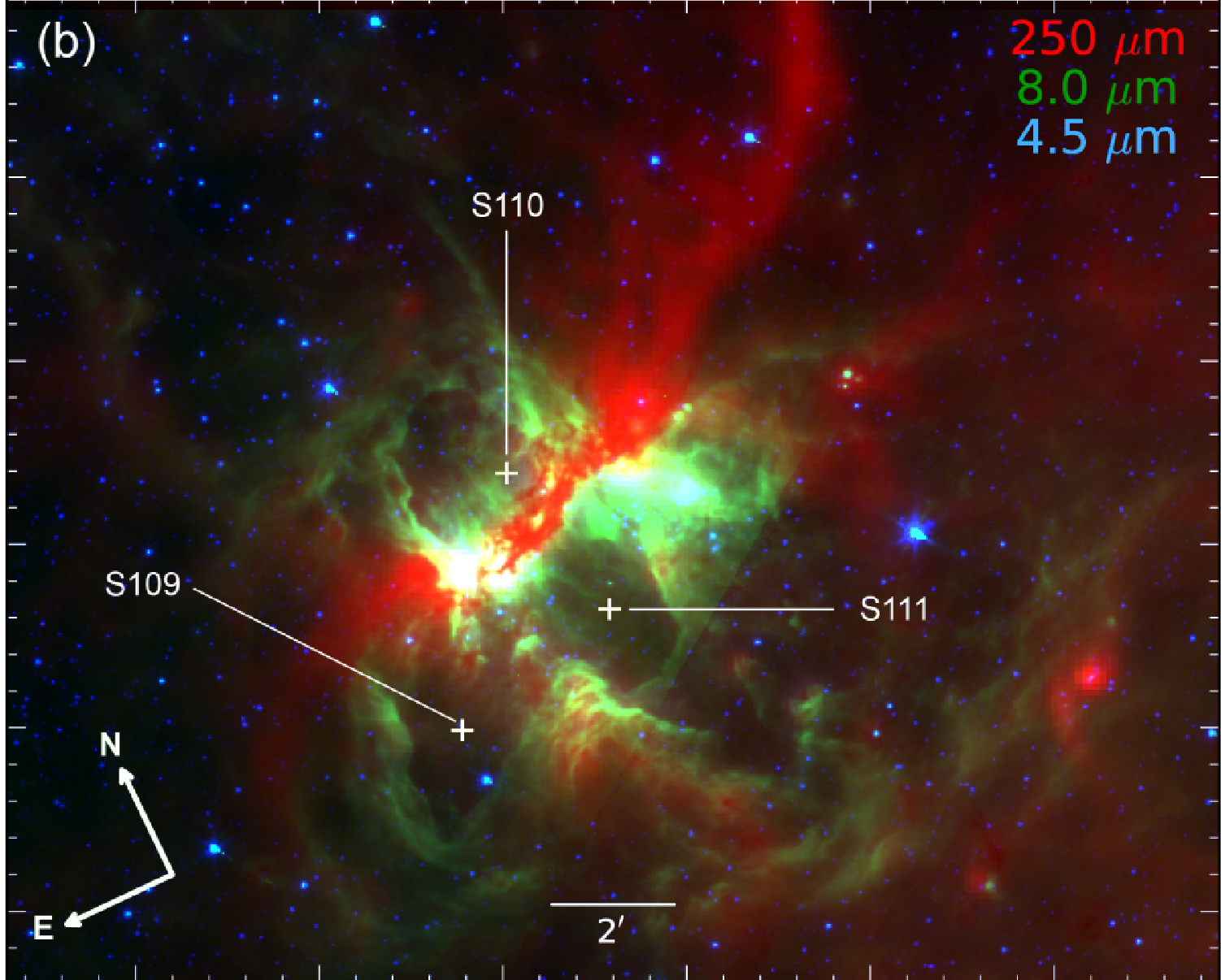}
\includegraphics[width=5.7cm]{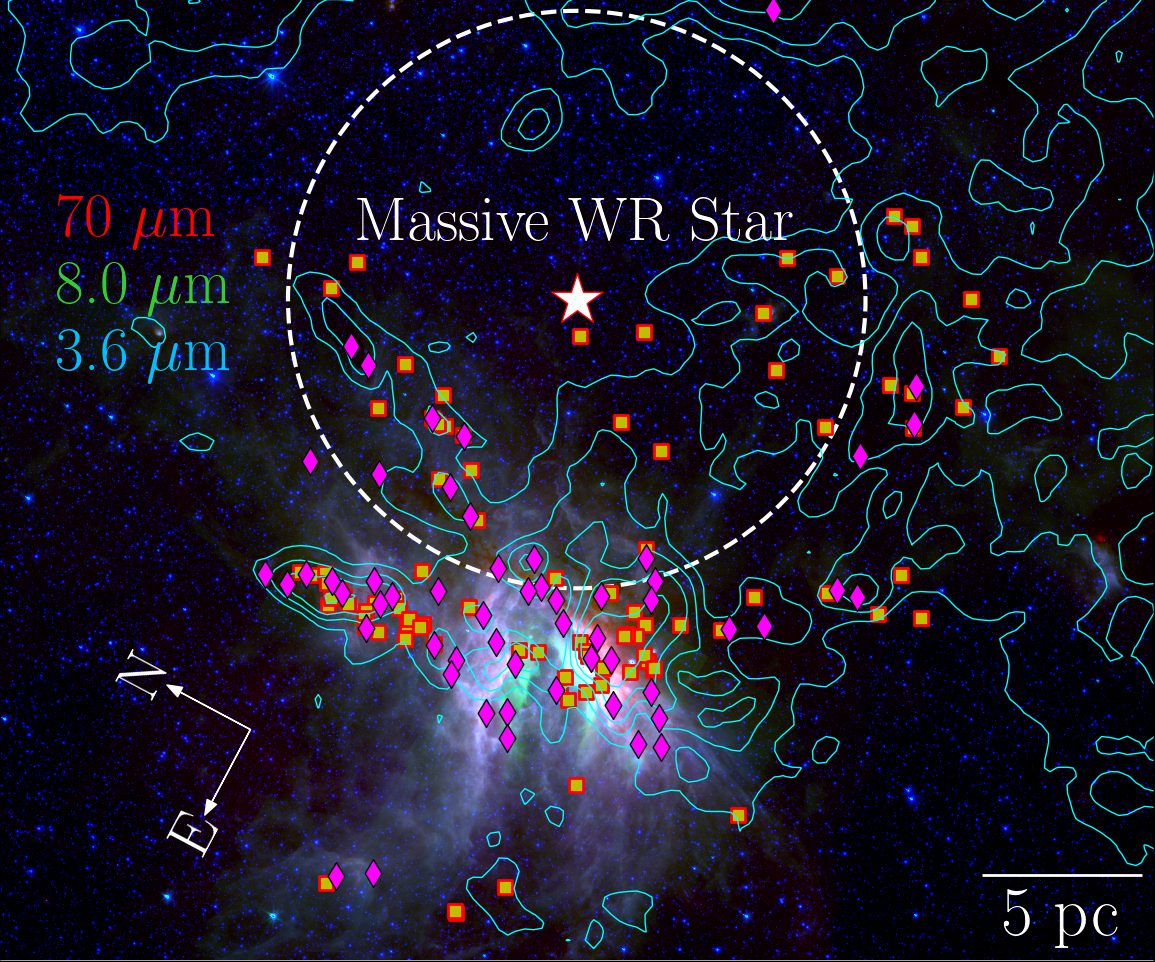}
\caption{(Left:) Schematic of different processes of triggered star formation around an \hii\ region -- 1:small-scale gravitational instabilities; 2: large-scale gravitational instabilities
 leading to the formation of high-mass fragments; 3: ionizing radiation acting on a turbulent medium; 4: radiation-driven compression of pre-existing dense clumps 
 \citep[adopted from][]{deharveng2010}. (Middle) Composite colour image of the bipolar nebula G316.80--00.05: red is for the Herschel 250 µm emission showing the cold filament at the waist of the nebula, green is for the 8 µm PAH emission, blue is for the 4.5 µm stellar emission. The three bubbles listed in the catalogue of mid-infrared bubbles by \citep{churchwell2006}  are also marked \citep{sam18}. (Right) A composite color (Red: Herschel 70\,\micron; Green: Spitzer 8\,\micron; Blue: Spitzer 3.6\,\micron) image of the G15.010-0.570 region. The selected sky-area includes a massive Wolf-Rayet (WR) star (asterisk), several cold dust clumps (magenta diamonds), Class 0 young stellar objects (red boxes), and one of the most active star-forming sites in the Galaxy, M17. Distribution of $^{13}$CO molecular gas is shown by cyan contours. Strong stellar winds from the WN7 type WR star have developed a molecular shell (white dashed circle), and star formation activity is noted around the expanding shell in the form of molecular clumps and Class 0/I YSOs. See \citep{baug2019} for more details on this study.
\label{fig:feedback}}
\end{figure*}

\subsection{Global status and open questions}

High resolution mid-infrared (8\,$\mu$m) observations from {\sl Spitzer} Space Telescope revealed the presence of bubble or shell-like structures around \hii\ regions \citep{churchwell2006, churchwell2007} arising from the thermal expansion of the \hii\ regions. In addition to the shells, a more homogeneous distribution of ionized gas pervaded by a web of filaments and large clouds are seen in giant massive star forming regions such the Cygnus X or W43. The ionizing radiation creates at the interface of \hii\ region/molecular clouds a rich variety of structures such as pillars, bright-rimmed clouds and globules.  \citet{deharveng2010} showed that the expansion of \hii\ regions could lead to different processes of triggered star formation, via different processes as shown in a schematic in Figure~\ref{fig:feedback}.  The collect and collapse (C\&C) mechanism of triggered star formation is typically observed around \hii\ regions/ bubbles \citep{zavagno2006, brand2011}. On the other hand, the radiation-driven implosion (RDI) mechanism is seen in pillars or bright rimmed clouds \citep{kessel2003, bisbas2011}. Figure\,\ref{fig:feedback} also shows observational evidence of the feedback in the form of a bipolar nebula and mid-infrared bubbles created by high-mass ZAMS stars in G316.80-00.05 \citep{sam18} as well as by Wolf-Rayet stars \citep{baug2019}. 

Several important aspects of feedback mechanism have been looked into from both theoretical and observational perspectives. It has been shown theoretically that stellar feedback could lead to inefficient star formation in giant molecular clouds \citep[GMCs;][]{hopkins2014, gatto2017, krumholz2019, grudic2022, bonne2022}, and even could potentially affect the maximum mass a massive star can achieve \citep{rosen2022}. In fact, a multi-wavelength study of \citet{olivier2021} on a large sample of more than hundred Galactic \hii\ region showed that feedback pressure is indeed sufficient to expel gas from the majority of the sources. Numerical simulations also showed the importance of wind feedback from massive stars in the context of star cluster formation. For
example, simulations of \citet{dale2013} revealed that the efficiency of wind-driven triggering to initiate star formation is low compared with the efficiency of the ionized gas.

The structures in the molecular cloud created by the ionizing radiation have long been observed in multiple wavelengths \citep{white1997,schneider2016,mookerjea2019, mookerjea2022} but details of how radiation and stellar wind impact the filaments and massive cloud ridges, and their impact on the subsequent filament/cloud evolution and star formation are topics of active research. In addition to isolated observations of PDRs associated with high-mass star-forming regions, recently a dedicated observational project (FEEDBACK) to study a sample of eleven high-mass star-forming regions primarily tracing the far-infrared emission from the PDRs using SOFIA \citep{schneider2020}. Preliminary results of the FEEDBACK project showed that massive stars have a variety of impacts and signatures on their surrounding environment, e.g., expanding wind-driven shell around a Wolf-Rayet star in RCW 49 region \citep{tiwari2021, tiwari2022}, sub-structured \hii\ region in NGC 7538 region that interacts with adjacent clouds and triggers formation of new-generation of stars \citep{beuther2022}. In a recent work using JWST near-and mid-infrared images extended intertwined/entangled sub-structures ($<$4500\,AU) consistent with instability in the dissociation front were observed towards the bubble wall of NGC 3324 \citep{Dewangan2023n3324}. More recently, study of PDRs in fine-structure lines of \CII\ at 158\,\micron\ and \OI\ at 63 and 145\,\micron\ using SOFIA have revealed the presence of a significant column of neutral gas not detected in CO \citep{guevara2020,goldsmith2021,mookerjea2021,mookerjea2023}. The origin of this gas is not yet completely understood.

\subsection{Status of the research in India}

In India, research devoted to the study of feedback of massive stars on the ambient ISM in the context of triggered star formation have primarily been based on observations using optical/near-infrared telescopes and the GMRT in the cm wavelengths \citep[see e.g.,][]{ojha2004a,ojha2004b, vig2007, tej2007, anandarao2008, chauhan2009, samal2010, sharma2012, dewangan2012,dewangan2016, dewangan2020a, mallick2015, dutta2015, baug2015, srdas2018, panwar2020, pandey2020}. To a limited extent studies of PDRs using far-infrared observations with the TIFR 100\,cm balloon-borne telescope, the Herschel Space Observatory and the airborne SOFIA observatory have also been pursued \citep{mookerjea2000,mookerjea2019,mookerjea2022,mookerjea2023}. A few key findings of these studies include: (i) observation and interpretation of molecular shells around massive O-type stars being created due to the ionization pressure, and the radiation pressure, (ii) identification of similar shells around the evolved Wolf-Rayet stars created by the impact of strong stellar winds, (iii) evidence of possible triggered star formation, based on over density of YSOs, in the high-density compressed shells created by expanding \hii\ regions, (iv) detection of non-thermal radio emission in an \hii\ region and (v) detection of large columns of atomic oxygen and C$^+$, the origin of which is still being explored.

\subsection{Key science questions and Goals}
Although substantial progress has been made in the field of feedback due to the massive stars, there still remains a few key questions in this area. These include but are not limited to i) What are the specific parameters that determine whether gas would be expelled or accumulated by the influence of massive stars? ii) What are the conditions that favor formation of the next generation of stars from the surviving parts of the original parental molecular cloud? iii) What is the effect of metallicity on the influence of stellar feedback? iv) Is the mass function in the clusters triggered by the influence of massive stars similar to the universal mass function? v) What is the heating efficiency of the FUV photons, and what determines that efficiency? vi) What is the contribution and effect of mechanical and radiative feedback on the surrounding clouds? Addressing these questions demands comprehensive parametric studies of a large number of Galactic massive star-forming regions and their natal environment. Successful execution of such projects would enable us to understand the differential nature of the stellar feedback and  proper conjecture/model of the stellar feedback would eventually help us to recover the star formation history in the galaxies as well in the Universe.

%
\section{Formation and evolution of binary and multiple star systems}


Stellar binarity and multiplicity is a ubiquitous feature across the H-R Diagram and may be critical to our understanding of nearly all aspects of Astronomy like the generation of gravitational wave to supernova rates to the occurrence of planetary systems. Even though the interdisciplinary importance of stellar multiplicity science is quite obvious yet their origin and evolution remains unclear. Numerical simulations point to two main formation routes for multiplicity: core fragmentation \citep{2004ApJ...600..769F} and disk fragmentation \citep{1989ApJ...347..959A}. In the core  fragmentation scenario, turbulent fluctuations within a bound core can develop multiple nonlinear density perturbations which when exceed the local Jeans mass could result in a collapse giving rise to multiple peaks. These peaks may subsequently become bound to form binary or multiple stellar system \citep[e.g., ][]{2010ApJ...725.1485O}. In disk fragmentation scenario \citep{1989ApJ...347..959A}, however, the  disks when subjected to sufficiently strong gravitational instability might fragment to form one or more companions. Binary separation provides one possible way of discerning which of these mechanisms are at work. While the disk fragmentation process favours formation of closer ($<$500 au) multiple systems, the turbulent fragmentation can produce multiple systems having initial separations $>$500 au \citep{2010ApJ...725.1485O}. Orientation of outflows from the stellar systems provides another possible way of distinguishing formation scenarios at work. Binaries forming through disk fragmentation is likely to have common angular momenta and, hence, aligned stellar spins, whereas binaries formed through turbulent fragmentation likely possess independent angular momentum direction and, thus, have randomly oriented spins \citep[e.g., ][]{2002A&A...387.1003M}.

To discern the formation mechanisms responsible for multiple systems, it is important to catch them at early protostellar stages of evolution in various environmental conditions to improve statistics on multiplicity, characterize the mass range, examine the separations closer to the au scales and study the limits of fragmentation and migration and then compare these observations with those of their evolved pre-main sequence counterparts, main sequence systems and with the multi-scale numerical simulations. Studies conducted at uniform sensitivities and resolutions towards the protostellar populations of Perseus \citep{2016ApJ...818...73T}, Ophiuchus \citep{2021ApJ...913..149E}, and Orion \citep{2020ApJ...905..162T, 2022ApJ...925...39T} using ALMA and Very Large Array (VLA) seems consistent despite environmental differences suggesting that formation mechanism of multiple systems in at least these three regions are probably similar. The companion separation distribution as a whole is double peaked inconsistent with that of solar-type field stars \citep{2022ApJ...925...39T}.
 
 Research in India on this rather important topic has been limited mainly due to the lack of facilities available within the country at sub-mm/mm wavelengths to observe multiplicity at protostellar phases. Even to characterize the multiplicity in evolved pre-main sequence systems, ground-based or space-based near- and mid-infrared facilities capable of producing high spatial resolution images are required. Consequently, most of the research works in India on this topic are limited to either statistical analysis \citep{2002A&A...387.1003M} or the study of individual sources \citep{2021MNRAS.501.1243A} using the data collected by the facilities abroad or through simulations and modelling \citep{2021Ap&SS.366...23M}.

 To improve the sample size of say multiplicity in protostars, it is essential to observe more populous and distant star forming regions. Since the spatial resolution reduces linearly with distance and the brightness of dust emission reduces as the square of the distance, we are reaching the limit of current capabilities. Though ALMA can achieve $\sim0.02^{\prime\prime}$ resolution at 1.3 mm, this emission will mainly detect optically thick dust in protostellar disks and not the close companions. Thus, a facility capable of $\sim0.01^{\prime\prime}$ angular resolution and $\sim10\times$ greater sensitivity operating between 3 mm to 2 cm is the pressing priority. If setting up of such a facility within India is challenging, it is prudent to join a consortium whenever such opportunities arise (like, ngVLA, FYST, AtLAST etc.).  While JWST will enable the detection and characterization of young multiple systems with close separations ($\sim0.1^{\prime\prime}$) via medium resolution spectroscopy, to determine the spectral types and in turn mass ratio and accretion diagnostics, high-resolution near-infrared spectroscopic instruments available with 10-m/30-m facility would be essential for long-term multi-object and multi-epoch campaigns to understand how multiple components are accreting matter and gaining mass over the time. 

 Though significant progress has been made over the past decade or so in the theoretical front, the efforts are still marred with poor resolution, lack of sufficient statistics and limited inclusion of local and global environmental effects in the treatment \citep{2022arXiv220310066O}. Therefore it is essential to adopt complex multi-physics treatment of the formation mechanisms of multiplicity that incorporates a large range of coupled physical processes, like self-gravity, supersonic turbulence, hydrodynamics, outflows, radiation, astrochemistry (disk chemistry), stellar feedback and B-fields. This requires use of advanced techniques like massive parallelism, multi-threading and advanced vector extensions. Today it is possible to get compute power rented from many public cloud providers including Amazon, Google and Microsoft which enable access to the latest generations of GPU devices, FPGAs and multi-core workstations to perform complex simulations.

\section{Planet-forming disks and Protoplanets}

\noindent
Star-forming regions are the ecosystems in which most stars and, therefore, most planets form. With the state-of-the-art observing capabilities of high sensitivity, high spatial and spectral resolution, such as ALMA, VLT, HST, and recently with JWST, the field of planetary system formation has undergone a transformational change.

\begin{figure*}
    \centering
    \includegraphics[width=14cm]{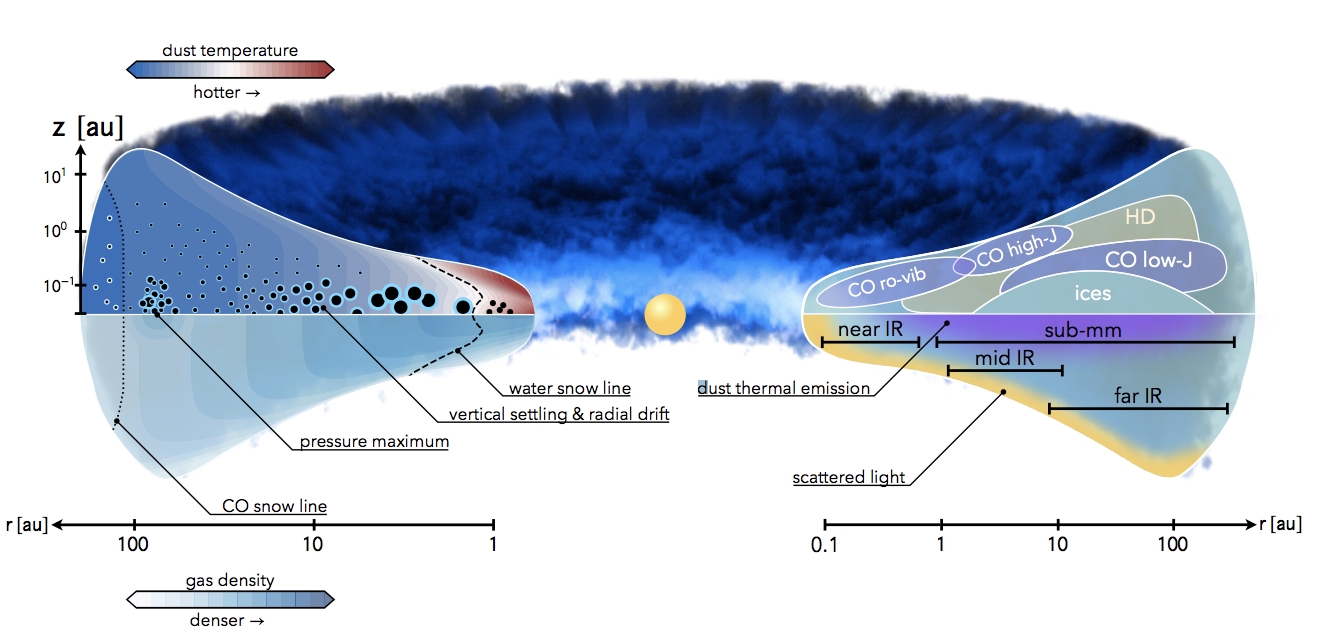}
    \caption{Sketch of a physical and chemical structure of a protoplanetary disk around a sun like star of age $\sim 1$--5\, Myr. Black circles are the distribution of dust particles of different sizes. The regions of main emissions coming from dust thermal, molecular and scattered light are  shown towards right \citep[adopted from][]{2023ASPC..534..501M}
    \label{fig:ppdisk}}
\end{figure*}

\noindent
With the advent of high-end observing capabilities,  highly sensitive complete surveys of star-forming regions involving imaging, spectroscopy, mm interferometry and polarimetry of young stars and their protoplanetary disks have become available. This provides an unprecedented statistical sample of young stellar objects with their physical parameters such as stellar masses, mass accretion rates, disk masses and radii, roughly within 300 pc from us. These observations also sample different evolutionary stages, ages, and environments.  The relations between stellar, accretion and disc properties, dependence of mass accretion rates on stellar masses, disc dust mass as a function of age are well sampled for the nearby regions \citep[see review by ][]{2023ASPC..534..539M}. In particular, with the ALMA,  the structure of accretion disks, protostellar multiplicity \citep[e.g., ][]{2022ApJ...925...39T}, and the chemical structure of pre-stellar cores \citep[e.g., ][]{2022ApJ...929...13C} have been established, though for limited samples (Fig.\,\ref{fig:ppdisk}). 
\noindent
In the context of theoretical model predictions, the traditional viscous evolutionary models, along with analytical descriptions of  MHD driven disk wind models, are widely acknowledged and tested to match the predictions with observations for the global disk properties (disk gas mass, radius, and accretion rate), their evolution in time and planet formation. However,  MHD disk wind models are yet to constrain large samples sufficiently (see the reviews by \citet{2023ASPC..534..539M}, \citet{2023ASPC..534..501M}). The evolution of protoplanetary discs through episodic accretion and stellar mass assembly has recently been a topic of discussion \citep{2023ASPC..534..355F}. The high-sensitivity and high-resolution imaging and spectroscopic capabilities of the JWST has led to detailed studies of the structure and chemistry of protoplanetary disks around both low- and high-mass stars \citep{vanDishoeck2023,Henning2024,Pontoppidan2024}. Discovery of new Herbig-Haro knots along with a clear identification of a single, isolated, low-mass protostar as the driving source supporting the idea of episodic accretion was also possible using the JWST mid- and near-infrared images \citep{Dewangan2024m16}.

\noindent
Indian researchers have primarily contributed to this area of research via disc evolution studies in diverse environments (mostly photometric analysis), disk fraction, accretion properties of Herbig Ae/Be type objects, and the variability analysis with the  time series photometric and spectroscopic observations using Indian and international observing facilities (e.g.,\citealt{2024ApJ...970...88P,2024arXiv240608261P,2023ApJ...954...82G, 2023MNRAS.520.1092L,2023MNRAS.524.5166N,2023MNRAS.521.5669C}). 

\subsection{Limitations in the current understanding}

Most of the above studies so far are on individual sources or small samples, limited within the solar neighborhood ($\sim$300 pc). However, most discs are expected to reside in much stronger UV environments, as well as high stellar density regions, which provide feedback in the form of external photoevaporation and dynamical interactions. Hence the above studies have the strong limitation of creating a biased view in sparse stellar groups where external photoevaporation or dynamical interaction is less effective. 

Understanding external feedback is, therefore, key to understanding the protoplanetary disc evolution followed by planet formation. In addition, theoretical models for the environmental influence on the chemical evolution of planetary discs and planets, and their predicted observational signatures are essential.  Ongoing and future JWST observations  of high UV environments will shed some light on disc chemistry from an observational perspective. Similarly, theoretical analysis on various steps to connect the correlation between protoplanetary and debris discs as well as those between disc properties and stellar mass are to be performed. Indeed, the current disc evolution models need to have a clear description of how these models scale with various stellar masses.  Hence, further constraints on the model parameters must be obtained to predict the scaling with both disc and stellar parameters. Finally, most of the observational effort till date has been on detailed studies of bright and massive disks, while the bulk of the disc population is heavily understudied. We lack deep CO surveys of large samples of discs and their outer radii in a representative number of star-forming regions. Metallicity is known to be another important parameter that decides the protoplanetary disk evolution followed by planet formation. We are yet to have any observational details and theoretical frameworks regarding the role of metallicity in disc and planet formation studies. A systematic analysis of protoplanetary disc evolution under low-metallicity environments of distant, outer Milky Way is the need of the hour to constrain their models. 

\subsection{Future Perspectives}

In order to study the protoplanetary discs of younger, embedded regions and evolving under extreme environmental conditions, high sensitive spectroscopic and imaging capabilities in the near-IR will be incredibly impactful, as they can provide better constraints on the initial conditions in star-forming clusters. Several multi-object spectroscopic facilities will be operational in the coming decade providing large scale surveys of young stellar objects in diverse environments. For example, the new 4-meter Multi-Object Spectroscopic Telescope (4MOST)  will be able to observe $\sim$2400 objects over a 4.2-degree hexagonal field of view (FOV) and provide both moderate (R$\approx$4000 - 7500) and high resolution (R$\approx$18,000 - 21,000) spectra. Another instrument, the Multi-object Optical and Near-IR spectrograph (MOONS) will be on the 8-m Very Large Telescope (VLT) of the European Southern Observatory (ESO). MOONS use the total 250 diameter FOV of the VLT, deploying 1001 fibers to obtain high-resolution spectra (R$\approx$20,000) with simultaneous optical and near-IR (0.6-1.8 $\mu$m) coverage. High-resolution coverage in the H-band will allow these observations to probe  embedded and crowded regions enabling similar spectral resolution in nearby regions to more distant and massive regions than Orion. Dedicated sensitive gas surveys of large samples of disks at moderate resolution are planned with ALMA and inner disk surveys using JWST. These two facilities complement each other: ALMA probes the outer and JWST the inner parts of disks. Future facilities in which India is actively involved, such as TMT, National Large Optical Telescope (NLOT), Indian Spectroscopic and Imaging Space Telescope (INSIST) etc will largely contribute to our understanding of this topic.


\section{Astrochemistry of ISM, protostars, and protoplanetary disks}


Astrochemistry deals with the study of the extent and cause of chemical enrichment of the ISM, in which over 240 different gas-phase and around 20 molecular species on the dust grain surface have so far been detected. Starting with the first detection of the diatomic molecule CH in the 1930s \citep{Swings1937, McKellar1940}, the complexity of the molecules discovered so far range from the diatomic H$_2$ to a 13-atom linear nitrile, HC$_{11}$N and bigger molecules such as 1-C$_{10}$H$_7$CN,2-C$_{10}$H$_7$CN and Fullerenes (C$_{60}$, C$_{60}^+$, C$_{70}$) (\cite{Tielens2008,McGuire2022}). The observational detections have also led to the development of numerical models considering chemistry in the gas-phase as well as gas-grain chemical networks to study the formation and abundance of the molecules \citep[][among others]{Watson1972,Herbst1973,Millar1975,Gould1963, Hasegawa1992, Garrod2008}. The development of astrochemistry has been supported by  laboratory experiments to measure reaction rates for gas-phase reactions (\cite{Smith2011} and reference therein)  and essential parameters associated with the gas-grain chemistry \citep[][among others]{Gerakines1996, Pirronello1997, Collings2003, Collings2004, Hama2013}, and references therein), which are crucial for the astrochemical models. 

\subsection{Global status \& Open Questions}

Enhanced observational capabilities in the sub-mm and millimeter wavelengths in the past decade has led to a significant improvement in our understanding of physical conditions and the chemical enrichment of Galactic and extra-galactic ISM \citep{Oberg2020, Jorgensen2020}. In particular the ALMA with its high sensitivity and high resolution spectroscopic capabilities has ushered in a boom in the discovery of new molecules, including Complex Organic Molecules (COMs). On the modelling front, state-of-the-art numerical simulation codes involving around 1200 species and 20,000 reactions are routinely used to study the formation of complex  organic molecules (COMs), including prebiotic molecules such as glycine \citep{Garrod2013, Agundez2013}. In addition, an attempt to understand isotopic fractionation and spin chemistry is underway, requiring even more extensive chemical networks. Furthermore, comparative studies between chemical complexity in solar system objects and proto-planetary disks are being investigated.  Besides, significant progress has been made to couple hydrodynamic simulations with chemical evolution, which will provide a better understanding. A significant progress in this regard are the public databases, such as KIDA, UMIST, NIST, etc., and the availability of numerical codes to the community.  Several state-of-the-art laboratories are built, which provide formation pathways and various physical parameters  such as sticking probability, adsorption and desorption energies, and activation barriers needed for astrochemical modelling. Laboratory formation of COMs, including simple amino acids  glycine- is also successfully studied \citep{Ioppolo2021}. Also, molecular spectroscopy databases involving laboratory measurement and computational molecular spectroscopy are created to aid the observation and identification of new molecules.

The availability of improved observations and models have also opened up many unresolved issues in understanding the chemistry of the molecules in the ISM, protostars and protoplanetary disks. Based on the presence of a large number of unidentified spectral lines, it is clear that the inventory of interstellar molecules is far from complete and this also requires accurate measurements of frequencies of many spectral lines in the laboratories.  Clear identification of chemical species resulting in the diffuse interstellar and aromatic infrared bands and detection of specific terrestrially abundant polycyclic aromatic hydrocarbons (PAHs) are among the highest in priority \citep{McGuire2021}. Molecules are often used as tracers (e.g., SiO, CH$_3$OH, N$_2$H$^+$, H$_2$CO, NH$_3$ etc.) of physical conditions such as radiation environment, density and temperature in numbers of environments, including ISM and star-forming regions.  Comparison of the observations with  numerical models to study the chemical and dynamical evolution of such regions together requires modification of the models using e.g., a multipoint Lagrangian scheme compared to traditional single point pseudo time-dependent chemical models. While,  modeling efforts are currently on to understand the formation of COMs and isoptopic fractionation of C, N and O, laboratory measurements and ab-initio calculations of  reaction rates, branching ratios, activation energies, and thermochemical data are needed for these efforts as well. Finally, this brings about the question of what the chemical diversity in the star-forming regions tell about the origin of Life on Earth? Answer to the question lies in the improved understanding of how chemical complexity changes as we move from diffuse to dense ISM, then to hot cores, to protoplanetary disks, to planetesimals, and finally, in the solar system objects.

\subsection{Research and status of the field in India}

Although “IAU Symposia 120: Astrochemistry” was organised in 1985 when the subject was in the early development phase, there was hardly any follow-up in India till early 2000s when a group was established to work on the modelling of the formation of complex molecules. Since then, activities primarily on astrochemical modeling followed closely by observations particularly in the post ALMA era have made significant contributions. These studies primarily focussed on the formation of molecules in diffuse and dense molecular clouds \citep{Chakrabarti2000, Das2008, Das2016,Iqbal2018a}, hot cores and hot corinos \citep{Majumdar2018, Acharyya2020, Sahu2020, Mondal2022, Srivastav2022, Acharyya2022} , protoplanetary disks as well as computational spectroscopy to generate spectra of COMs and PAHs \citep{Pathak2008,Buragohain2020,Vats2022, Ahmad2020, Singh2021a,  Maurya2015}. Using the Herschel space observatory and SOFIA, in several papers Indians have led the work on the detection of linear carbonchain molecules CCC (and its isotopes) towards high mass star-forming regions \citep{Mookerjea2010,Giesen2020}. These work have reported for the first time in the literature evidence of Warm Carbon Chain Chemistry in high-mass star forming regions \citep{Mookerjea2012} and detection of C$^{13}$CC and $^{13}$CCC in the interstellar medium. In the recent times, Indian authors have constrained physical processes at various stages of star and planet formation through chemical modelling of sub-mm spectral lines \citep{Das2020, Gorai2020,
Majumdar2016, Majumdar2018, Sahu2020, Sahu2021}. Laboratory experiments and theoretical 
studies for spectral identification and ice composition and to derive parameters for surface chemistry have also been undertaken \citep{Acharyya2014,Bhuin2014,Sivaraman2015a, Methikkalam2016,Pavithraa2017a,Chakraborty2019, Ghosh2019, Vishwakarma2021}. 

\subsection{Key science questions, future goals etc., from an Indian perspective.}

It will not be an overstatement to say that the holy grail of Astrochemistry lies in understanding the processes which led to the formation of a habitable planet like ours and the origin of life on it. The key questions which lead up to an understanding of the origin of life include: (i) How, when, and where are molecules produced and excited? (ii) What do they tell us about temperatures, densities, gas masses, ionization rates, radiation fields, and dynamics of the clouds? (iii) How are the molecules cycled through the various phases of stellar evolution, from birth to death? (iv) How far does the molecular chemical complexity go? More specifically, the studies focus on water and organic molecules which need to be supplied to terrestrial worlds like our own to provide the essential compounds required for the origin of life. These molecules form initially during the earliest stages of stellar birth, are supplied by collapse to the planet-forming disk predominantly as ice, and may undergo significant processing during this collapse and within large planetesimals that are heated via radioactive decay. Thus, the key questions in the field are centered around the formation mechanism of Complex Organic Molecules (COMs) in the low-temperature interstellar medium. The dominant mechanisms at such low temperatures are still a matter of dispute, with both gas-phase and granular processes, occurring on and in ice mantles, thought to play a role. Efforts of the Indian groups are focussed on the detection of molecules at high resolution in the star forming cores and protoplanetary disks to constrain the indigenously developed numerical models. These efforts are being complemented by laboratory measurements and theoretical studies of surface chemistry on interstellar ice.



\section{Summary \& Recommendation}
\begin{figure*}[!h]
\centering
\resizebox{\textwidth}{10cm}{\includegraphics{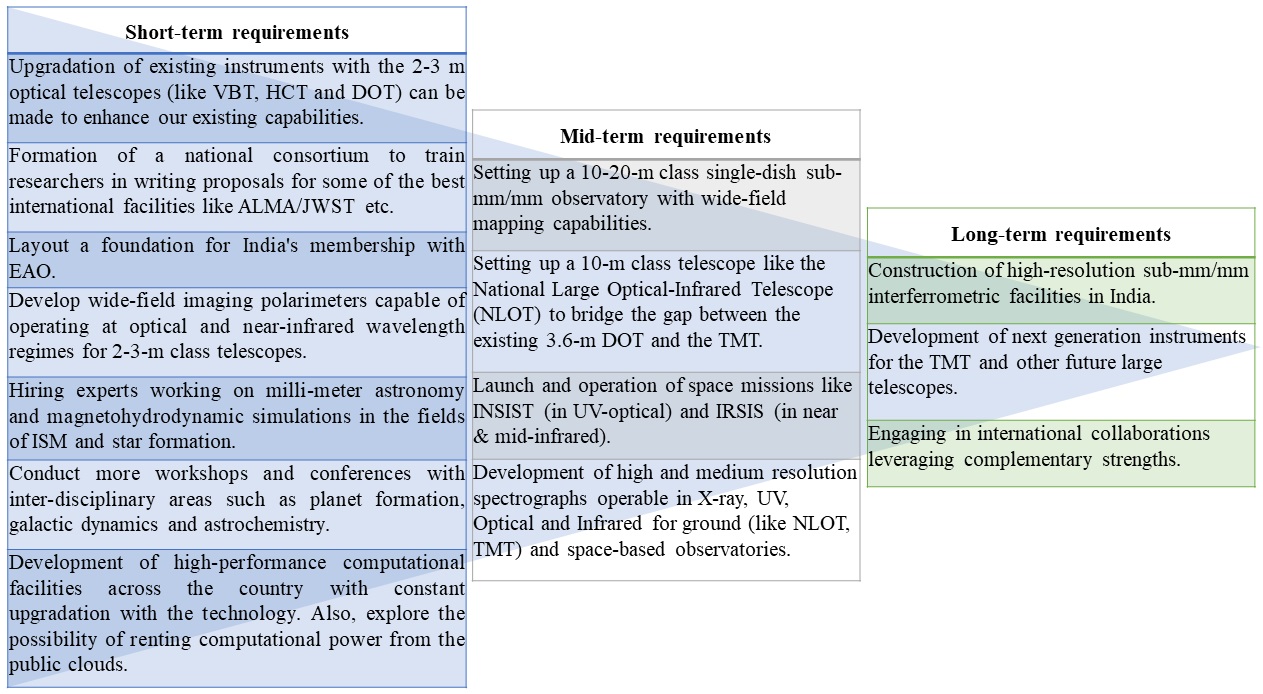}}
\caption{Recommendations and suggestions from the ISM and Star Formation group.}\label{fig:reco}
\end{figure*}

The ISM and star formation community in India has about 100 -- 150 active researchers, including Ph.D. students, postdocs, and faculty members working in various institutions and universities. The community gathers annually in different parts of the country, where the members hold discussion/brainstorming sessions to showcase the results and summarize the status, challenges, and prospects. A significant fraction of the science done by the community is based on archival data and individual collaborative access to state-of-the-art-telescopes. However in order to be able to address  the problems pertaining to star formation, large observational programs needing guaranteed access to state-of-the-art telescopes are required.  The community thus recommends development of the state-of-the-art telescopes, back-end instruments, computational facilities and access to international facilities. Some of the key recommendations are listed below. These recommendations/requirements are categorized into short- (5 years), mid- (10 years) and long-term (15-20 years). A brief summary of the recommendations is presented in Figure\,\ref{fig:reco}.

\underline{\bf Short-term requirements:}\\
\(\bullet\) The instruments available in many of the existing Indian facilities were commissioned long ago and, hence, lacked modern capabilities to cater to the growing community. Therefore we recommend upgradation of existing instruments with the 2-3-m optical telescopes (like VBT, HCT and DOT) to enhance the existing capabilities. As modern instruments are complex and expensive, developing future instruments in collaborations across multiple institutions is prudent. Pooling resources and sharing infrastructure can result in cost savings and operational efficiencies for participating institutions. Joining the small telescopes with international programs such as LCOGT to conduct coordinated time series observations will enhance their productivity.\\
\(\bullet\) ALMA is the best facility in the mm domain in terms of the resolution and sensitivity for studying star formation at the finest scale, thus, access to this facility would allow the Indian community to be competitive in the field for years to come. Therefore it is recommend to build a national consortium to write proposals in the best international facilities like ALMA/JWST etc. This will allow the researchers in the country to get trained with the data from the best facilities and also to become competitive in the field for years to come. \\
\(\bullet\) East Asian Observatory (EAO) has been allowing Indian PIs to submit proposals for JCMT under the Expanding Partner Program to basically encourage astronomers from new JCMT's partners. Under this program, PIs from Thailand, Malaysia, Vietnam, Indonesia and India requesting $<$ 15 hours will be automatically approved upon satisfying the technical feasibility and scientific merit. Some of the Indian PIs have already utilized this opportunity. However, it is needed to lay down a foundation for India's membership with EAO to utilize the capabilities of JCMT, SMA, etc. This will allow Indian PIs to get access to sub-mm/mm facilities something that is seriously lacking.\\
\(\bullet\) Development of wide-field imaging polarimeters capable of operating at optical and near-IR wavelength regimes for 2-3-m class telescopes should be of high priority to construct 2-D maps of B-fields in ISM.\\
\(\bullet\) Hiring of experts working in sub-/millimetre wavelengths and magnetohydrodynamic simulations in the field of star formation as faculties to perform high-end simulations and also to train students, postdocs and younger faculties in data reduction and interpretation and to use upcoming facilities such as ngVLA.\\
\(\bullet\) Efforts should be made to conduct more workshops and conferences with inter-disciplinary areas such as planet formation, galactic dynamics, and astrochemistry to enhance our understanding of the complex problems of star formation.\\
\(\bullet\) Like observatories, development of a few high-performance computational facilities across the country is also essential for understanding the kinematics, morphology, and chemical evolution of molecular gas, effectively through simulations and also to be used for data intensive computing. Today it is possible to get compute power rented from many public cloud providers including Amazon, Google and Microsoft which enable access to the latest generations of GPU devices, FPGAs and multi-core workstations to perform complex simulations. Though this requirement is listed under short-term category, according to the technological advancement over the period, our capability should also be upgraded in timely manner.

\underline{\bf Mid-term requirements:}\\
\(\bullet\) Currently, there are no active sub-mm/mm facilities in the country, which has hampered our progress significantly in the field of star-formation research. Thus, in the coming decade, we must invest in setting up a 10-20-m class single-dish sub-mm/mm observatory with wide-field mapping capabilities. Mapping our Galaxy in multiple dust continuum bands as well as with a few key dense gas tracer lines such as NH$_3$, HCN, HCO$^+$, N$_2$H$^+$, CS and H$_3$CN will allow us to take a major leap in a few key areas of star formation. It will also complement the data of the other existing sub-mm/mm facilities.  Access to single dish millimetre facilities is also essential, not only to understand large-scale star-formation processes, but also to complement interferometric data, particularly for the extended gas structures. Access to the above facilities is also essential to be ready to make use of upcoming high-cadence future facilities such as ngVLA that would be beneficial to study star and planet formation. Sub-mm/mm telescope with polarimetric capability would allow us to trace the B-field geometry of molecular clouds.\\
 \(\bullet\) Installation of a 10-m class telescope like NLOT is essential to bridge the gap between the existing 3.6-m DOT and the upcoming TMT. The new observatory should have AO assisted instruments, integral field spectroscopy and multi-object spectroscopy. The 10-m class telescope should be used as a test bed for the prototyping of new technology instruments and to train man power for the upcoming and future bigger telescopes. Till we have our own 10-m telescope, it is highly recommended to purchase time in some of the existing 10-m class telescopes such as Subaru, GTC, SALT, etc. We can also join consortium that are open for partnerships through in-kind contributions.\\
 \(\bullet\) Medium- and high-resolution spectroscopy in UV, optical and near- and mid-infrared wavelengths from the ground and space are important for the studies of the ISM and star formation. Thus development of multi-object spectrographs in these wavelengths for the upcoming ground-based facilities (e.g. NLOT, TMT) is required. 
It is also important that the proposed space-based missions such as Indian Spectroscopic and Imaging Space Telescope (INSIST) and the  InfraRed Spectroscopic Imaging Surveyor (IRSIS) both of which have already received initial funding come to fruition. The INSIST is a 1-m class telescope capable of delivering images at a high-spatial resolution of $0.1^{\prime\prime}-0.2^{\prime\prime}$ at wavelengths ranging from 0.15-0.55 $\mu$m. It also has spectroscopic capability (low- to medium-resolution) in multi-object mode. The UV region provides one of the direct methods of measuring accretion in young stars as a result of material plunging onto stellar surface creating UV radiation due to shock. The region also has some of the prominent spectral lines (like O I $\lambda1304$ \AA, C IV $\lambda$1550 \AA, He II $\lambda$1640 \AA, C II $\lambda$2330 \AA, Mg II $\lambda$2798 \AA) that are of interest to the star formation research community. The IRSIS, on the other hand, is a medium size telescope of 30-cm aperture diameter capable of collecting and focusing infrared radiation from astronomical objects in the sky and feed to an array of micro-lenses coupled to infrared optical-fiber bundles. The primary science goals of IRSIS, proposed by the TIFR group include detection of several spectral lines \& features due to the Interstellar Gas and Dust components of the ISM of our Galaxy. 

\underline{\bf Long-term requirements:}\\
\(\bullet\) Construction of high-resolution sub-mm/mm interferrometric facility in India, for conducting spectroscopic, dust continuum, and polarization observations of dense molecular clouds. A facility capable of $\sim0.01^{\prime\prime}$ angular resolution and $\sim10\times$ greater sensitivity operating between 3 mm to 2 cm is also essential to identify close companions to stars at their earliest stages of star formation.\\
\(\bullet\) Development of second generation instruments for the TMT capable of performing high-spatial resolution imaging and spectroscopy of young stars to detect close multiple components, study accretion \& outflows and other properties towards the fainter end of the stellar mass spectrum in distant massive star forming regions residing in diverse environments.\\
\(\bullet\) Explore possibilities of engaging in international collaborations leveraging complementary strengths. Combining each partner's unique capabilities and assets will allow us to put more complex and powerful UV/infrared/sub-mm facilities in space to address some of the forefront problems in ISM and star formation.


\bibliography{ms_16Aug24}

\clearpage
\label{lastpage}

\end{document}